\documentclass[12pt]{iopart}
\usepackage{iopams}
\usepackage{graphicx}
\usepackage{subfigure}
\usepackage{color}

\begin{document}

%%%%%%%%%%%
%% Title %%
%%%%%%%%%%%

% Note: The title in [ ] will appear at the top of each page
%	The title in { } is the proper title of the article

\topical[Band engineering in dilute nitride and bismide semiconductor lasers]{Band engineering in dilute nitride and bismide semiconductor lasers}

%%%%%%%%%%%%%%%%%%%%%%%%%%%%%%
%% Authors and affiliations %%
%%%%%%%%%%%%%%%%%%%%%%%%%%%%%%

\author{C. A. Broderick$^{1,2}$, M. Usman$^{1}$, S. J. Sweeney$^{3}$ and E. P. O'Reilly$^{1,2}$}

\address{$^{1}$ Tyndall National Institute, Lee Maltings, Dyke Parade, Cork, Ireland \\
	 $^{2}$ Department of Physics, University College Cork, Cork, Ireland \\
	 $^{3}$ Advanced Technology Institute and Department of Physics, University of Surrey, Guildford, Surrey, GU2 7XH, United Kingdom}

\ead{eoin.oreilly@tyndall.ie} % Email address of corresponding author

%%%%%%%%%%%%%%
%% Abstract %%
%%%%%%%%%%%%%%

\begin{abstract}
Highly mismatched semiconductor alloys such as GaN$_{x}$As$_{1-x}$ and GaBi$_{x}$As$_{1-x}$  have several novel electronic properties, including a rapid reduction in energy gap with increasing $x$ and also, for GaBiAs, a strong increase in spin-orbit-splitting energy with increasing Bi composition. We review here the electronic structure of such alloys and their consequences for ideal lasers. We then describe the substantial progress made in the demonstration of actual GaInNAs telecomm lasers. These have characteristics comparable to conventional InP-based devices. This includes a strong Auger contribution to the threshold current. We show, however, that the large spin-orbit-splitting energy in GaBiAs and GaBiNAs could lead to the suppression of the dominant Auger recombination loss mechanism, finally opening the route to efficient temperature-stable telecomm and longer wavelength lasers with significantly reduced power consumption.  
\end{abstract}

%%%%%%%%%%%%%%%%%%%%%%%%%%%%%%%%%%%%%%
%% Specify PACS numbers and journal %%
%%%%%%%%%%%%%%%%%%%%%%%%%%%%%%%%%%%%%%

% \pacs{xx} % PACS numbers
\submitto{\SST} % Specify journal (SST = Semicond. Sci. Technol.)

\maketitle

%%%%%%%%%%%%%%%%%%%%%%%%%%%%%
%% Section I: Introduction %%
%%%%%%%%%%%%%%%%%%%%%%%%%%%%%

\section{Introduction}
\label{sec:introduction}

The exponential growth in optical telecommunications and the internet has been underpinned by the development of semiconductor lasers emitting at 1.3 $\mu$m and at 1.55 $\mu$m, the wavelengths at which respectively dispersion is zero and losses are minimised in standard optical fibres \cite{Zory_book,Kapon_books}. The lasers designed to operate at these wavelengths are based primarily on the growth of quaternary InGaAsP and InGaAlAs alloy structures on InP substrates. Despite their widespread application, there are several significant drawbacks associated with these devices, mostly associated with the constraints of growing on InP substrates. The development of telecomm lasers grown on GaAs substrates could bring several advantages. Firstly, because GaAs is a more robust material, growth can be carried out on larger substrates. Secondly, better optical confinement can be achieved in GaAs heterostructures, because of the larger refractive index difference between GaAs and AlGaAs compared to that between InP and the quaternary alloys. Vertical cavity surface emitting lasers can therefore be grown monolithically on GaAs, but only with great difficulty on InP. In addition, because AlGaAs has a considerably larger energy gap than InP and can be grown lattice-matched on GaAs, it should be possible to achieve much better electrical confinement in a telecomm laser based on GaAs. A further issue with 1.3 and 1.55 $\mu$m semiconductor lasers is that their threshold current and the optical cavity losses tend to increase strongly with increasing temperature, due largely to a combination of two intrinsic loss mechanisms, Auger recombination \cite{Higashi_JSTQE_1999,Sweeney_PSSb_1999,Silver_JQE_1998} and intervalence band absorption \cite{Adams_JJAP_1980}. Telecomm lasers then need to be operated on a thermo-electric cooler for many applications, significantly increasing the overall energy budget associated with their operation.

Considerable advantage could therefore be gained if high quality and reliable telecomm lasers could be developed on GaAs. This is very difficult to achieve using conventional quantum well structures. Highly efficient lasers using strained InGaAs quantum well structures are very well established for emission around 1 $\mu$m, but too much In is needed and there is therefore too large a lattice mismatch relative to GaAs to achieve reliable lasers emitting at 1.3 $\mu$m and beyond \cite{Coleman_Zory_Chpt8}. There was therefore considerable interest generated when it was shown that replacing a small fraction of As atoms by N in Ga(In)As leads to a rapid reduction in energy gap, with the gap decreasing by about 150 meV for $x = 1$\% in GaN$_{x}$As$_{1-x}$ \cite{Kondow_JJAP_1996,Kondow_JSTQE_1997}. This opens the possibility to achieve longer wavelength emission on a GaAs substrate, supporting the demonstration of 1.3 $\mu$m edge-emitting and vertical cavity lasers \cite{Steinle_EL_2001,Livshits_EL_36,Riechert_SST_2002}, as well as 1.5 $\mu$m edge-emitting devices \cite{Bank_EL_2006,Bank_JQE_2004}.

The electronic structure of dilute nitride alloys is of considerable interest from a theoretical perspective, both to understand the origins of the rapid reduction in energy gap and also to understand the consequences of that reduction for laser action. It is well established that replacing a single As atom by N in GaAs introduces a resonant defect state above the conduction band edge \cite{Wolford_Proc_ICPS_1984}. A major breakthrough was achieved for dilute nitride alloys with the demonstration by Walukiewicz and co-workers that the reduction in energy gap in Ga(In)N$_{x}$As$_{1-x}$ is due to a band-anticrossing interaction between the conduction band edge and higher-lying localised nitrogen resonant states \cite{Shan_PRL_1999}. 

Given the significant differences in the conduction band structure of GaInNAs compared to conventional III-V semiconductors, it is important to elucidate the influence of N not only on the electronic structure but also on the gain characteristics of ideal dilute nitride lasers. We have shown \cite{Tomic_JSTQE_2003,Tomic_Photon_Tech_Lett_2003} and describe below that the incorporation of N degrades properties such as differential gain compared to that of the best GaInAs/GaAs structures. Nevertheless, the overall characteristics of an ideal GaInNAs/GaAs laser are still expected to be at least as good as those of conventional InP-based telecomm lasers.

Having established the characteristics of ideal structures, we then turn to consider actual GaInNAs lasers operating in the telecomm wavelength band between 1.3 and 1.5 $\mu$m. We have shown experimentally that the threshold current of GaInNAs lasers is dominated by non-radiative recombination processes, including both defect-related and Auger recombination \cite{Fehse_JSTQE_2002}. Theoretical calculations \cite{Andreev_APL_2004} and experimental analysis \cite{Fehse_JSTQE_2002} show that Auger recombination is a major loss mechanism in GaInNAs- and in InP-based 1.3 and 1.5 $\mu$m lasers. In the dominant CHSH Auger process, a \textbf{C}onduction electron and a \textbf{H}eavy hole recombine across the energy gap but, instead of emitting a photon, they excite into the \textbf{S}pin-split-off band a \textbf{H}ole from near the valence band maximum \cite{Sweeney_PSSb_1999}. 

CHSH Auger recombination is the dominant intrinsic loss mechanism that ultimately limits all current telecomm lasers. Because Auger recombination involves three carriers, its contribution to the threshold current density increases (using the Boltzmann approximation) as $n_{\rm{th}}^{3}$, where $n_{\rm{th}}$ is the threshold carrier density. The rapid increase of $n_{\rm{th}}^{3}$ with temperature then explains the strong temperature sensitivity of the threshold current in semiconductor diode lasers at telecomm and longer wavelengths. In addition to Auger recombination, inter-valence band absorption (IVBA) also degrades laser performance. In the IVBA process, a photon is re-absorbed by an electron in the spin-split-off band which is excited to an empty (hole) state near the top of the valence band. Like the CHSH Auger process, IVBA is therefore also sensitive to the spin-split-off energy. IVBA causes $n_{\rm{th}}$ to increase superlinearly with temperature \cite{Adams_SST_1987}, further exacerbating Auger recombination and additionally degrading the differential quantum efficiency above threshold. The combination of Auger recombination and IVBA causes the output power of telecomm lasers to be strongly temperature sensitive over the typical operating temperature range \cite{Sweeney_IEEE_LEOS_2003}. 

It would therefore be highly beneficial if these loss mechanisms could be eliminated from telecomm lasers. This cannot be achieved using conventional III-V alloys based on (GaAlIn)(NPAsSb), where the energy gap $E_{g}$ always exceeds the spin-orbit splitting energy $\Delta_{\scalebox{0.7}{\rm{SO}}}$ in materials that emit at 1.3 and 1.5 $\mu$m. It has recently been demonstrated that replacing As by Bi in GaBi$_{x}$As$_{1-x}$ can give rise to a strong decrease in the band gap as well as a strong increase in the spin-orbit-splitting energy, leading to a $\Delta_{\scalebox{0.7}{\rm{SO}}} > E_{g}$ regime in the alloy for $x \gtrsim 10$\% \cite{Sweeney_ICTON_2011,Usman_PRB_2011}. Just as the introduction of N gives the possibility to engineer the conduction band structure, the introduction of Bi then provides an opportunity to engineer the valence band structure potentially providing a route to produce telecomm and longer wavelength lasers with suppressed losses \cite{Sweeney_ICTON_2011,Sweeney_patent_2010}. 

This review aims to provide an overview of the electronic structure of highly mismatched semiconductor alloys and of the consequences and potential benefits of such alloys for optoelectronic applications. In Section 2 we present an overview of the band-anticrossing model used to describe the conduction band structure of dilute nitride alloys. A brief discussion of the effective masses and band offsets used in calculations on GaInNAs/GaAs material systems is given in Section 3. In Section 4, we discuss the effects of nitrogen incorporation on gain and loss mechanisms in ideal GaInNAs lasers, followed by a comparison between ideal and actual device behaviour. Having established through this comparison that Auger recombination remains a significant loss mechanism in all telecomm lasers to date, we then present in Section 5 an overview of the potential benefits of dilute bismide alloys for future, high efficiency photonic devices. Finally we summarise our conclusions in Section 6.

%%%%%%%%%%%%%%%%%%%%%%%%%%%%%%%%%%%%%%%%%%%%%%%%%%%%%%%%%%%%%%%%%%%%%%%%%%%%%%%%%%%%%
%% Section II: Band-anticrossing model of dilute nitride conduction band structure %%
%%%%%%%%%%%%%%%%%%%%%%%%%%%%%%%%%%%%%%%%%%%%%%%%%%%%%%%%%%%%%%%%%%%%%%%%%%%%%%%%%%%%%

\section{Band-anticrossing model of dilute nitride conduction band structure}
\label{sec:conduction_band_structure}

It is well-established that when a single N atom replaces an As atom in GaAs, it forms a resonant defect level above the conduction band edge of GaAs \cite{Wolford_Proc_ICPS_1984,Liu_APL_1990}. This defect level arises because of the large difference in electronegativity and atomic size between N and As \cite{Vogl_Adv_Electron_Phys_1984,Hjalmarson_PRL_1980}. A major breakthrough was achieved for dilute nitride alloys with the demonstration by Walukiewicz and co-workers using hydrostatic pressure techniques that the reduction in energy gap in Ga(In)N$_{x}$As$_{1-x}$ is due to a band-anticrossing (BAC) interaction between the conduction band edge and higher-lying localised nitrogen resonant states \cite{Shan_PRL_1999}.

The BAC model explains the extreme band gap bowing observed in In$_{y}$Ga$_{1-y}$N$_{x}$As$_{1-x}$ in terms of an interaction between two levels, one at energy $E_{c}$ associated with the extended conduction band edge (CBE) state $\psi_{c0}$ of the GaInAs matrix and the other at energy $E_{\rm{N}}$ associated with the localised N impurity states $\psi_{\rm{N}}$, with the two states linked by a matrix element $V_{\rm{Nc}}$ describing the interaction between them \cite{Shan_PRL_1999}. The conduction band (CB) dispersion of Ga(In)N$_{x}$As$_{1-x}$ is then given in the BAC model by the lower eigenvalue of the determinant

\begin{equation}
  \label{eq:2_band_BAC}
  \left| \begin{array}{cc}
      E_{\rm{N}} - E & V_{\rm{Nc}} \\
      V_{\rm{Nc}} &  E_{\rm{c}} + \frac{ \hbar^{2} k^{2} }{ 2 m^{*}_{c} } - E \end{array} \right|
\end{equation}
where $m^{*}_{c}$ is an appropriately chosen CBE effective mass for the Ga(In)As host matrix material \cite{Tomic_PRB_2004}. From Eq.~\ref{eq:2_band_BAC}, the CBE energy, $E_{-}$, is then given by

\begin{equation}
  \label{eq:e_minus}
  E_{-} = \frac{ E_{\rm{N}} + E_{c} }{ 2 } - \sqrt{ \left( \frac{ E_{\rm{N}} - E_{c} }{ 2 } \right)^{2} + \left( V_{\rm{Nc}} \right)^{2} } .
\end{equation}
The alloy CBE wavefunction $\psi_{-}$ can be found from Eq.~\ref{eq:2_band_BAC} as $\psi_{-} = \alpha_{c} \psi_{c0} + \alpha_{N} \psi_{N}$. The fractional $\Gamma$ character of $\psi_{-}$ is given by $f_{\Gamma} = \vert \alpha_{c} \vert^{2}$, with $f_{\Gamma}$ then providing a useful measure of how much the N-related states in the alloy perturb the CBE wavefunction.

\begin{figure}[t!]
  \centering
  \hspace{-1.0cm}
  \subfigure[]{ \vspace{1.0cm} \includegraphics[scale=0.32]{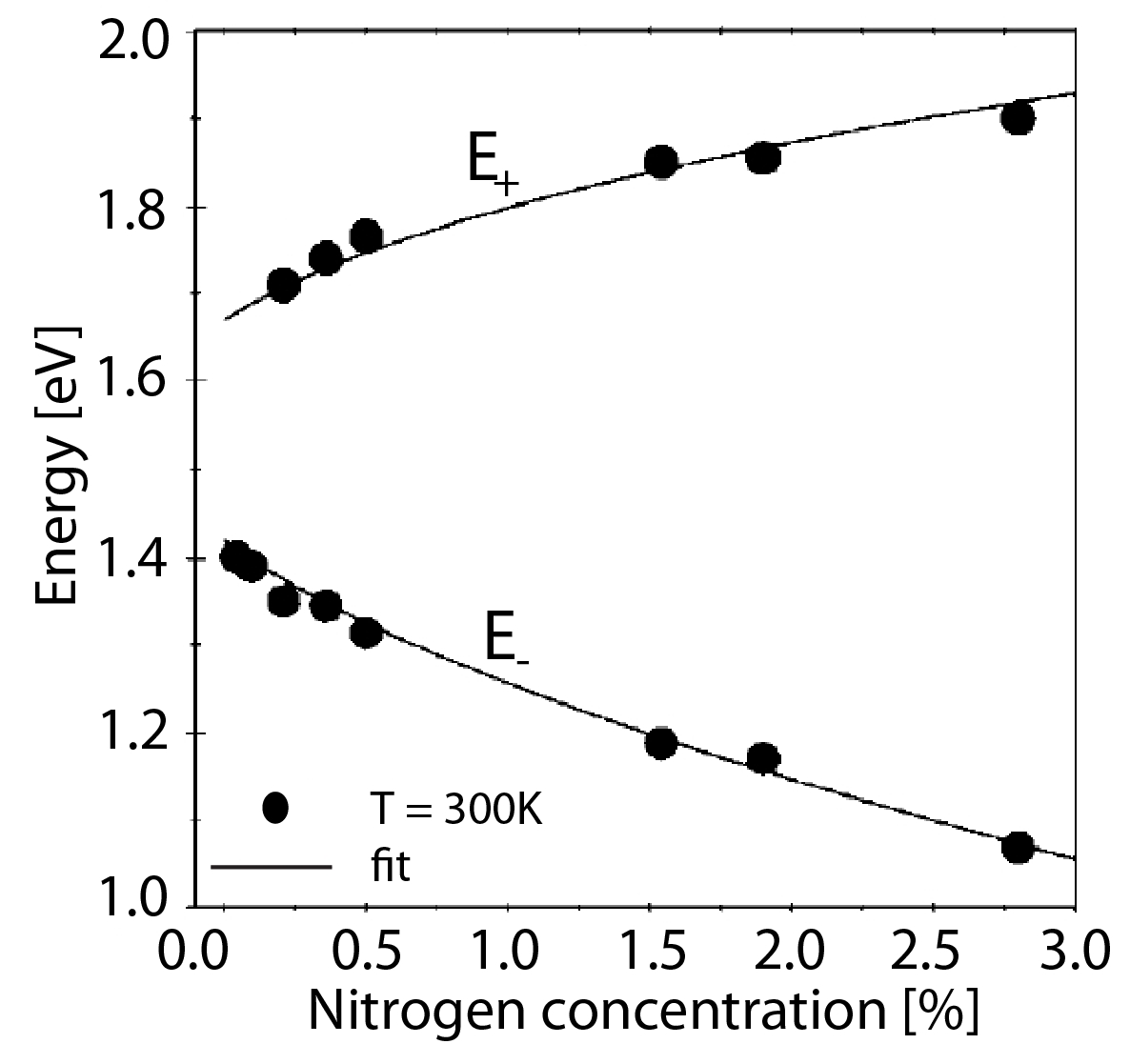} \label{fig:GaNAs_PR} }
  \hspace{-0.5cm}
  \subfigure[]{ \includegraphics[scale=0.32]{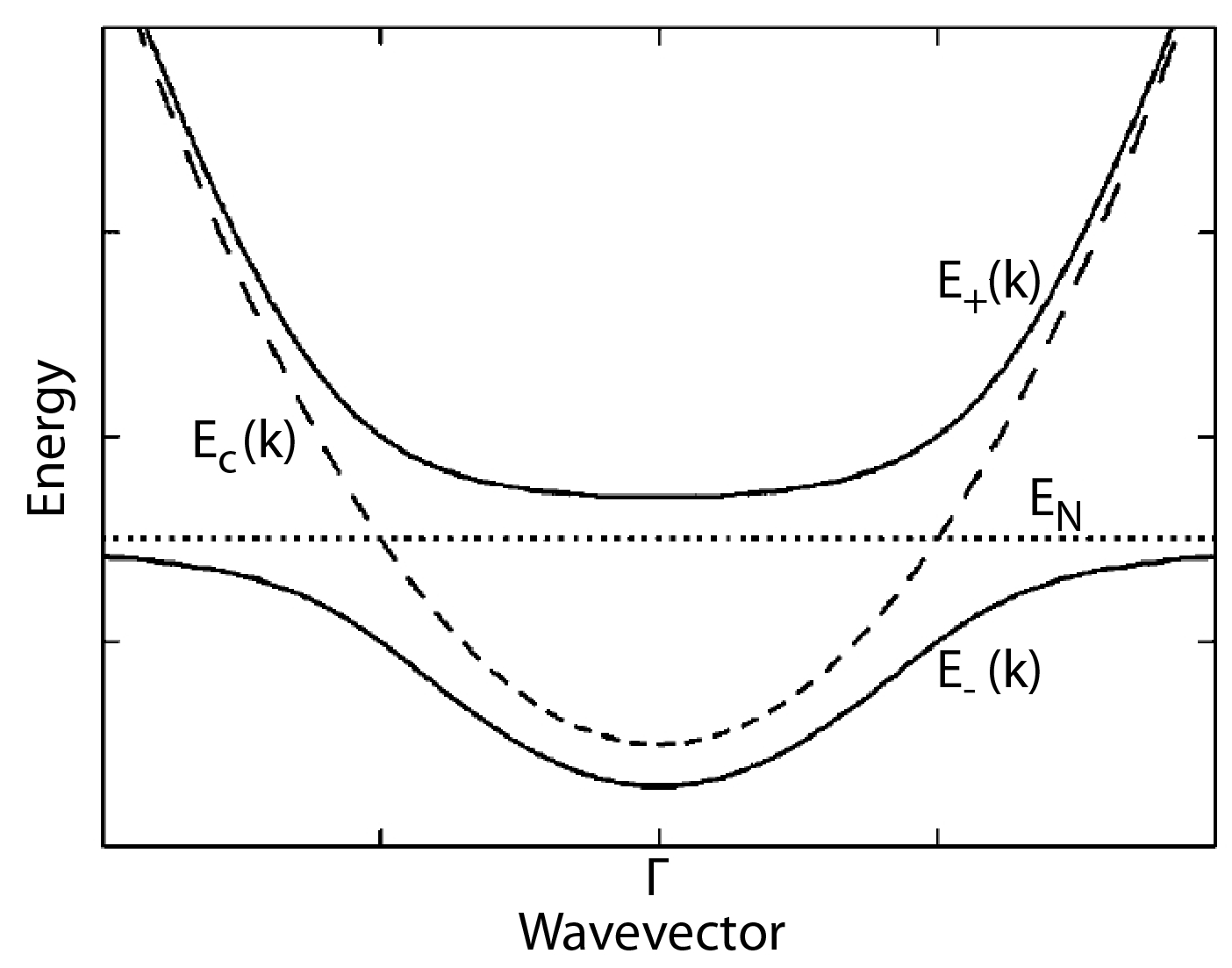} \label{fig:GaNAs_BAC_dispersion} }
  \caption{(a) Points: Variation of the upper and lower nitrogen related bands, $E_{+}$ and $E_{-}$, obtained from room temperature photo-reflectance measurements on GaN$_{x}$As$_{1-x}$. Solid lines: Calculated variation of $E_{+}$ and $E_{-}$ using the 2-band BAC model of Eq.~\ref{eq:2_band_BAC}. (b) Solid lines show the conduction band dispersion calculated using Eq.~\ref{eq:2_band_BAC}. Dashed line: Host matrix conduction band dispersion; dotted line: N resonant defect energy level. ((a) and (b) reproduced from Ref.~\cite{Reilly_SST_2009}.)}
\end{figure}

A resonant feature associated with the upper eigenvalue, $E_{+}$, has been observed in photo-reflectance (PR) measurements \cite{Shan_PRL_1999,Perkins_PRL_1999,Klar_APL_2000}, appearing in GaN$_{x}$As$_{1-x}$ for $x \gtrsim$ 0.2\% and remaining a relatively sharp feature until $x \sim$ 3\%, beyond which composition it broadens and weakens, when the resonant state becomes degenerate with the GaAs $L$-related conduction band levels \cite{Lindsay_PhysicaE_2004}. The data points in Figure~\ref{fig:GaNAs_PR} show the measured variation of the $E_{-}$ and $E_{+}$ energy levels with N composition $x$ at room temperature in GaN$_{x}$As$_{1-x}$. These measured values are very well fitted using the BAC model, as illustrated by the solid lines in Figure~\ref{fig:GaNAs_PR}, for which we use $E_{\rm{N}} = 1.67$ eV, $E_{c} = 1.42 - 3.5x$ eV and $V_{\rm{Nc}} = \beta x^{\frac{1}{2}}$, with $\beta =$ 2.3 eV \cite{Klar_SST_2002}.

% Figure: BAC behaviour in Ga(In)NAs

The BAC interaction not only reduces the energy gap but is also predicted to lead to an increased electron effective mass at the CB minimum and to a strongly nonparabolic CB dispersion. Evidence for this dispersion is provided by photo-reflectance measurements of GaNAs quantum well (QW) samples, where the strong band nonparabolicity is required to account for the QW excited state transition energies across a wide range of samples and as a function of hydrostatic pressure \cite{Klar_SST_2002,Tomic_PRB_2004}.

Despite the wide success of the 2-level BAC model, it should be acknowledged that there are several sets of experimental data which it fails to explain. For example, it significantly underestimates the electron relative effective mass, $m^{*}_{e}$, in GaN$_{x}$As$_{1-x}$ for $x >$ 0.1\% \cite{Buyanova_PRB_2000,Hai_APL_2000,Baldassarri_PRB_2003,Masia_APL_2003,Reilly_SST_2009}. The effective mass has been determined using a range of different techniques, with a consistent trend emerging of unexpectedly large relative mass values in GaN$_{x}$As$_{1-x}$, such as $m^{*}_{e} =$ 0.13, 0.15 and even 0.19 for $x =$ 0.1 \cite{Masia_APL_2003}, 1.6 \cite{Baldassarri_PRB_2003} and 2.0\% \cite{Hai_APL_2000}, respectively.

Detailed calculations have shown that these effects arise due to the formation of N pairs and clusters in the material, which can generate defect states lying close to the CBE of GaN$_{x}$As$_{1-x}$ \cite{Reilly_SST_2009,Lindsay_PRL_2004,Kent_PRB_2001}. Hybridization between these localised states and the host matrix CBE strongly perturbs the electronic structure, with the strength of the hybridization determined by the relative separation in energy between the N-related defect states and the CBE \cite{Reilly_SST_2009,Lindsay_PRL_2004}.

In InAs the localised states related to N pairs and higher order complexes lie significantly above the InAs CBE, so that replacing Ga by In to form In$_{y}$Ga$_{1-y}$N$_{x}$As$_{1-x}$ shifts the CBE away from the N cluster state distribution \cite{Reilly_SST_2009}. This significantly reduces the effects of N cluster states on the CBE, meaning that the BAC model has been largely successful when applied to In$_{y}$Ga$_{1-y}$N$_{x}$As$_{1-x}$ alloys, accounting well for the observed BAC-like energy gaps and effective masses in these alloys \cite{Reilly_SST_2009,Tomic_PRB_2005}.

The electronic structure of conventional semiconductor lasers is typically analysed using an 8-band \textbf{k}$\cdot$\textbf{p} model, including the conduction band, heavy-hole, light-hole and spin-split-off bands \cite{Meney_PRB_1994,Yan_Voon_book}. This can be expanded to a 10-band \textbf{k}$\cdot$\textbf{p} model for GaInNAs, obtained by adding two additional spin-degenerate nitrogen-related states to the 8-band \textbf{k}$\cdot$\textbf{p} model \cite{Tomic_JSTQE_2003,Lindsay_SSE_2003}. This model has provided valuable insight into the optoelectronic properties of GaInNAs alloys, having been successfully used to accurately model, for example, the measured optical transitions \cite{Klar_SST_2002} and electron effective masses \cite{Tomic_PRB_2004,Tomic_PRB_2005} in GaInNAs quantum wells (QWs), as well as gain and loss mechanisms in 1.3 $\mu$m GaInNAs QW lasers \cite{Tomic_JSTQE_2003,Fehse_JSTQE_2002}.

%%%%%%%%%%%%%%%%%%%%%%%%%%%%%%%%%%%%%%%%%%%%%%%%%%%%%%%%%%%%%%%%%
%% Band gap, band offsets and effective masses in GaInNAs/GaAs %%
%%%%%%%%%%%%%%%%%%%%%%%%%%%%%%%%%%%%%%%%%%%%%%%%%%%%%%%%%%%%%%%%%

\section{Band gap, band offsets and effective masses in GaInNAs}
\label{sec:band_offsets_and_effective_masses}

\begin{figure}[t!]
  \centering
  \includegraphics[scale = 0.36]{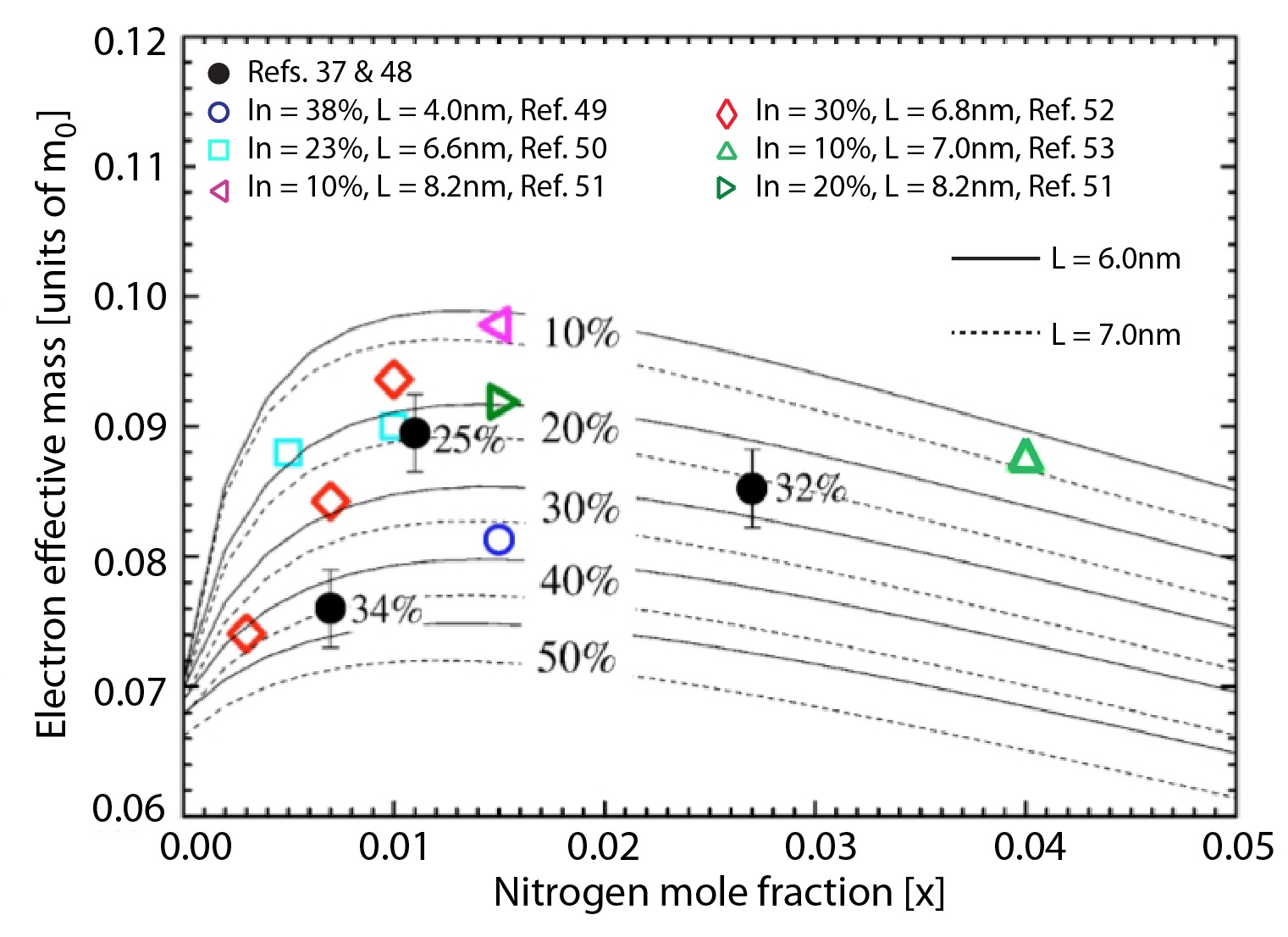}
  \caption{Solid and dotted lines: Calculated variation of the in-plane CBE electron effective mass, $m^{*}_{e}$ as a function of N composition, $x$, in a series of In$_{y}$Ga$_{1-y}$N$_{x}$As$_{1-x}$ QWs, of width 6 nm (solid lines) and 7 nm (dotted lines) and for In composition increasing from $y = 10$\% to 50\% in steps of 10\%. Data points: Experimental $m^{*}_{e}$ values taken from the indicated sources~\cite{Baldassarri_PRB_2003, Polimeni_SSE_2003, Hetterich_APL_2000, Duboz_APL_2002, Heroux_JAP_2002, Pan_APL_2001, Gass_APL_2004}. (Figure adapted from Ref.~\cite{Tomic_PRB_2005}.)}
  \label{fig:GaInNAs_effective_masses}
\end{figure}

A range of studies have successfully interpreted the band structure of GaInNAs alloys using the BAC model. As an example, Figure~\ref{fig:GaInNAs_effective_masses} compares the measured CBE effective mass in GaInNAs with that predicted using the  2-level BAC model. In all cases the effective mass is increased compared to that in GaAs, due to the BAC interaction. It can be seen that there is generally good agreement between the theoretical and experimental data, reflecting that there are few or no cluster states near to the CBE in this alloy \cite{Reilly_SST_2009}.

It is impossible to provide a unique set of BAC parameters for GaInNAs alloys. This is because the band structure depends on the local environment about each N atom. The energy $E_{\rm{N}}$ of the isolated N defect energy levels shifts down in energy as Ga neighbours are replaced by In neighbours \cite{Reilly_SST_2009}. The magnitude of the BAC interaction, $V_{\rm{Nc}}$ also decreases with increasing number of In neighbours, because the In-N equilibrium bond length is larger than the Ga-N one, leading to a reduced distortion about a N site with increasing number of In neighbours. The energy gap of GaInNAs alloys is typically found to blue shift after brief thermal annealing \cite{Tournie_APL_2002,Jouhti_APL_2003}. This is attributed to an increase in the average number of In neighbours on annealing \cite{Jouhti_APL_2003}. BAC parameter sets have been developed to describe annealed GaInNAs \cite{Tomic_JSTQE_2003,Healy_JQE_2006}. Although there are differences between the various reported sets, these differences do not affect the overall conclusions discussed below.

% Figure: Theory vs. expt. for GaInNAs effective masses

There was considerable initial uncertainty as to the choice of conduction and valence band (VB) offsets to use in GaInNAs/GaAs QW structures, with the choice of band offset values also impacting the gain characteristics. Because of the strong band-anticrossing interaction, the CB offset and the band offset ratio vary strongly with N composition in Ga(In)NAs/GaAs heterostructures and therefore it is most useful to work with the VB offset. We assume that the introduction of N to form In$_{y}$Ga$_{1-y}$N$_{x}$As$_{1-x}$ leads to a linear variation in VB edge energy in the unstrained alloy, which we describe by the parameter $\kappa$ such that $E_{\rm{v}} (x,y) =  E_{\rm{v}0} (y) + \kappa x$, where $x$ is the nitrogen fraction. How does introduction of N affect the VB offset? This is difficult to measure by optical spectroscopy, because of the number of free parameters available when fitting to ground and excited state transition energies, including the nitrogen energy, $E_{\rm{N}}$,  BAC coupling parameter, $\beta x^{\frac{1}{2}}$ and the band offset $\kappa$ parameter. As a consequence, a wide variety of offset values were presented in the literature. These range from values that give a type I structure for a In$_{y}$Ga$_{1-y}$N$_{x}$As$_{1-x}$/GaAs QW \cite{Tomic_PRB_2004,Buyanova_PRB_2000,Klar_PSSb_2001} to those that favour a type II structure \cite{Sun_APL_2000}. Values of $\kappa$ used in the literature have ranged from -1.87 eV \cite{Sun_APL_2000} to 3.5 eV \cite{Tomic_PRB_2004}. Galluppi et al. \cite{Galluppi_APL_2005} undertook a systematic study of the HH VB offset in a series of GaInNAs samples with In concentrations up to 34\% and for a range of N concentrations up to and above 2\%. They found a small decrease in the VB offset with increasing N concentration, predominantly because of the reduction in the HH-LH splitting due to the reduction in compressive strain in the InGaNAs QWs due to N incorporation. A similar reduction was also observed for smaller In concentrations (0 -- 20\%) by Heroux and coworkers \cite{Heroux_JAP_2002}, who also attributed the slight decrease in VBO to a reduction in the net compressive strain. A small value of $\kappa \approx$ 0.2 eV has also been deduced from tight-binding calculations \cite{Lindsay_SSC_1999}. Based on these experimental and theoretical results, we therefore recommend a value of $\kappa =$ 0 to model the band structure of GaInNAs/GaAs heterostructures.

With the choice of $\kappa=0$, there is then a large conduction band offset and comparatively small valence band offset in GaInNAs. This has led to the concern that the small valence band offset may lead to leakage currents contributing to the total threshold current in GaInNAs lasers \cite{Tansu_JAP_2005,Carrere_APL_2005}. However, self-consistent calculations show that because the electrons are strongly confined in GaInNAs/GaAs QW structures, the high electron density provides an additional electrostatic contribution to the valence band confinement potential, thereby reducing the impact of the small valence band offset \cite{Healy_JQE_2006}.

%%%%%%%%%%%%%%%%%%%%%%%%%%%%%%%%%%%%%%%%%%%%%%%%%%%%%%%%%%%%%%%%%%%%%%%%%%%
%% Section IV: Gain and recombination processes in dilute nitride lasers %%
%%%%%%%%%%%%%%%%%%%%%%%%%%%%%%%%%%%%%%%%%%%%%%%%%%%%%%%%%%%%%%%%%%%%%%%%%%%

\section{Gain and loss mechanisms in dilute nitride lasers}
\label{sec:gain_and_recombination}

%%%%%%%%%%%%%%%%%%%%%%%%%%%%%%%%%%%%%%%%%%%%%%%%%%%%%%%%%%%%%%%%%%%%%%%%%%%%%%%%%%
%% Influence of N on gain and radiative current density in ideal GaInNAs lasers %%
%%%%%%%%%%%%%%%%%%%%%%%%%%%%%%%%%%%%%%%%%%%%%%%%%%%%%%%%%%%%%%%%%%%%%%%%%%%%%%%%%%

\subsection{Influence of N on gain and radiative current density in ideal GaInNAs lasers}
\label{sec:Influence_of_N_on_gain}

The major driver towards the development of dilute nitride III-V compounds was the realization by Kondow and coworkers \cite{Kondow_JJAP_1996,Kondow_JSTQE_1997} that these alloys offer a viable route to achieve GaAs-based quantum well (QW) laser diodes emitting in the 1.3 and 1.5 $\mu$m optical windows. These GaInNAs/GaAs QW structures also provide the benefits associated with compressively strained devices \cite{Adams_Elec_Lett_1986,Reilly_SST_1989}. In addition, the inclusion of N in the GaInNAs layers increases the CB offset, leading to improved electron confinement when compared with conventional InGaAsP 1.3 $\mu$m lasers \cite{Kondow_JJAP_1996,Kondow_JSTQE_1997}. Both edge-emitting and vertical-cavity (VCSEL) laser structures have been reported with impressive characteristics, as discussed further below. 

Given the significant difference between the CB structure of GaInNAs and conventional III-V semiconductors, it is important not only to elucidate the influence of N on the electronic structure and gain characteristics of ideal dilute nitride lasers, but also to apply that understanding to the experimental analysis of GaInNAs lasers operating in the telecomm wavelength band between 1.3 and 1.5 $\mu$m.

We present here a theoretical overview of the consequences of the electronic structure of dilute nitride alloys for laser emission. We summarise the influence of nitrogen incorporation on the calculated dipole matrix elements, electron effective mass, gain and differential gain versus carrier density and radiative current density, as well as the effect of N on the primary loss mechanisms of defect-related and Auger recombination. We then compare the theoretically estimated laser characteristics of GaInNAs-based and conventional InGaAsP-based 1.3 $\mu$m lasers.

% Figure: GaInNAs/GaAs SQW calculated (a) band structure and (b) transition matrix elements

\begin{figure}[t!]
  \centering
  \hspace{-0.5cm}
  \subfigure[]{ \includegraphics[scale=0.39]{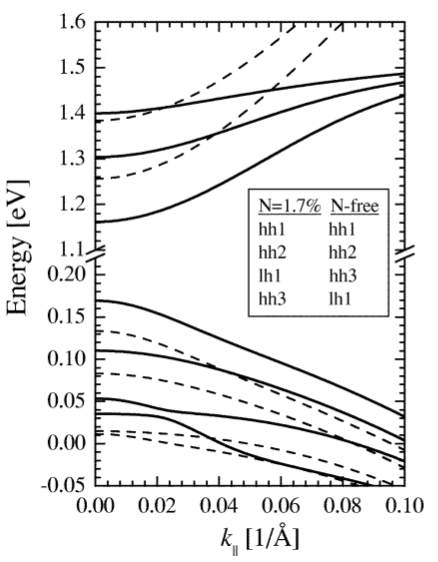} \label{fig:GaInNAs_QW_band_structure} }
  \hspace{-0.5cm}
  \subfigure[]{ \includegraphics[scale=0.35]{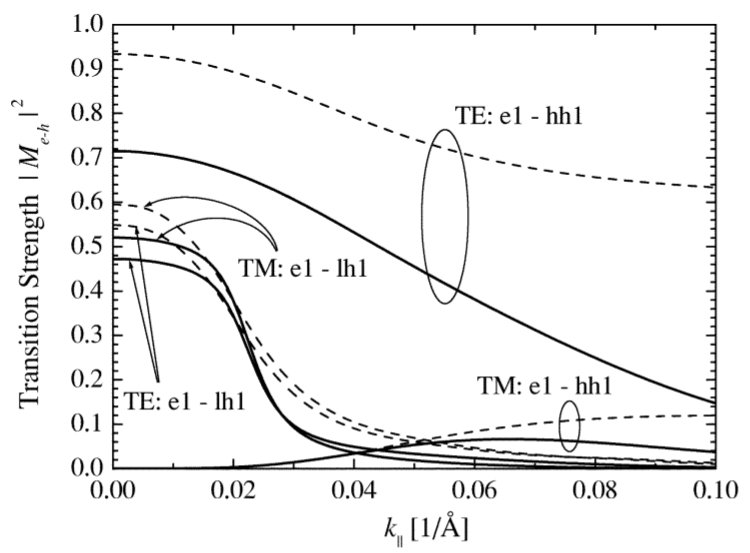} \label{fig:GaInNAs_QW_transition_strengths} }
  \caption{(a) Solid lines: Band structure of an In$_{0.36}$Ga$_{0.64}$N$_{0.017}$As$_{0.983}$/GaAs SQW laser device calculated from a 10-band \textbf{k}$\cdot$\textbf{p} Hamiltonian which was parametrised to fit the experimentally observed QW transitions at $k_{\parallel} = 0$ \cite{Tomic_JSTQE_2003}. Dashed lines: Calculated band structure of the equivalent N-free SQW device described in the text. (b) Calculated transition matrix elements for the same N-containing (solid lines) and N-free (dashed lines) SQW laser devices. ((a) and (b) reproduced from Ref.~\cite{Tomic_JSTQE_2003}.)}
\end{figure}

The solid lines in Figure~\ref{fig:GaInNAs_QW_band_structure} show the calculated band dispersion for an ideal 7 nm wide In$_{0.36}$Ga$_{0.64}$N$_{0.02}$As$_{0.98}$/GaAs quantum well, calculated using a 10-band \textbf{k}$\cdot$\textbf{p} Hamiltonian which includes the CB, heavy-hole (HH), light-hole (LH) and spin-split-off (SO) states from the conventional 8-band \textbf{k}$\cdot$\textbf{p} model, as well as two additional spin-degenerate N-related states \cite{Tomic_JSTQE_2003}. The dashed line shows the calculated band dispersion of the equivalent N-free 7 nm wide In$_{0.36}$Ga$_{0.64}$As/GaAs quantum well structure. An increased band edge effective mass and a strong nonparabolicity are clearly visible in the CB dispersion in the GaInNAs QW, due to the interaction between the lowest conduction states and the N resonant levels.

The coupling between the N level and the CBE modifies the CB wavefunctions and reduces the interband optical transition matrix element $\vert M_{\rm{e-h}} \vert^{2}$ compared to a conventional N-free alloy. Figure~\ref{fig:GaInNAs_QW_transition_strengths} shows the calculated variation as a function of in-plane wavevector $k_{\parallel}$ for the TE and TM interband matrix elements linking the first confined electron ($e1$) with heavy-hole ($hh1$) and light-hole ($lh1$) states. The band edge, zone-centre, TE matrix element is calculated to decrease by $\sim$ 30\% due to the incorporation of N. The N-CB coupling therefore leads to an increased conduction band edge effective mass. We calculate that the band edge density-of-states (DOS) effective mass in the first conduction band increases to $0.060m_{0}$  in the structure with 2\% N compared to a value of $0.046m_{0}$ in the N-free case.

% Figure: Calculated variation of GaInNAs/GaAs SQW peak material gain as a function of (a) carrier density and (b) radiative current density

\begin{figure}[t!]
  \centering
  \hspace{-1.0cm}
  \subfigure[]{ \vspace{1.0cm} \includegraphics[scale=0.32]{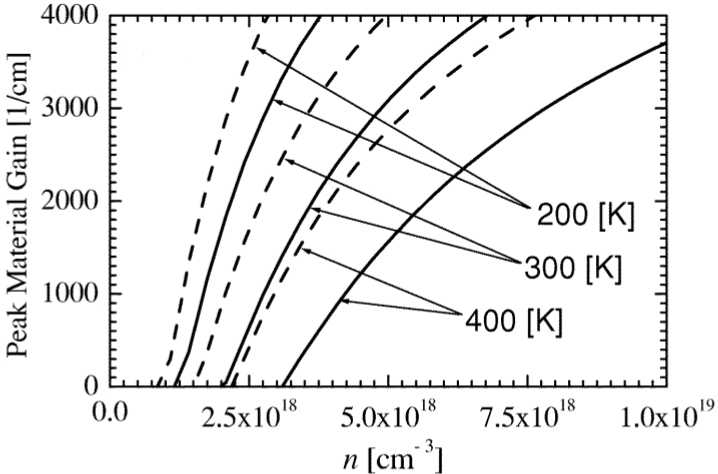} \label{fig:GaInNAs_QW_gain_n} }
  \hspace{-0.5cm}
  \subfigure[]{ \includegraphics[scale=0.32]{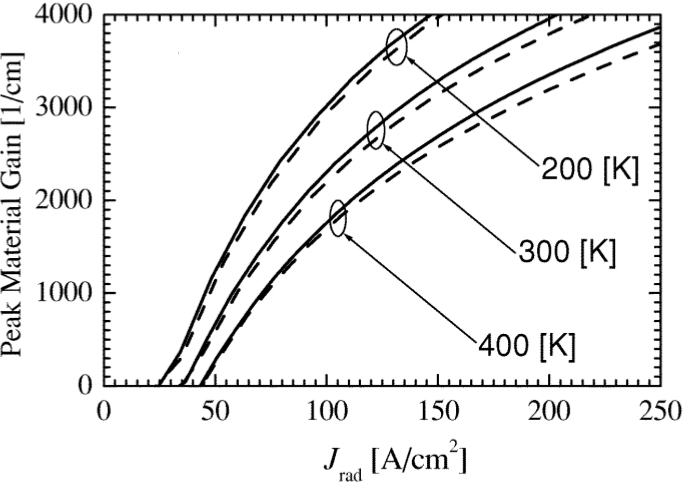} \label{fig:GaInNAs_QW_gain_Jrad} }
  \caption{Calculated variation of the peak material gain as a function of (a) carrier density and (b) radiative current density, in the same N-containing (solid lines) and N-free (dashed lines) SQW laser device of Figures~\ref{fig:GaInNAs_QW_band_structure} and~\ref{fig:GaInNAs_QW_transition_strengths}. ((a) and (b) reproduced from Ref.~\cite{Tomic_JSTQE_2003}.)}
\end{figure}

The reduction in $\vert M_{\rm{e-h}} \vert^{2}$ and consequent increase in the conduction band edge effective mass, $m^{*}_{e}$, causes the product $m^{*}_{e} \times \vert M_{\rm{e-h}} \vert^{2}$ to stay approximately constant, as would be expected from \textbf{k}$\cdot$\textbf{p} theory: the dominant contribution to the bulk conduction band inverse effective mass, $m^{* \; -1}_{e}$, is directly proportional to $\vert M_{\rm{e-h}} \vert^{2}$. The band structure and matrix elements presented in Figures~\ref{fig:GaInNAs_QW_band_structure} and~\ref{fig:GaInNAs_QW_transition_strengths} were used to calculate the variation of material gain with temperature $T$ and as a function of carrier density $n$ and of radiative current density $J_{\rm{rad}}$ both in an In$_{0.36}$Ga$_{0.64}$N$_{0.02}$As$_{0.98}$/GaAs QW structure and in an equivalent N-free structure \cite{Tomic_PhysicaE_2002}.  The increase in the conduction band effective mass leads to an increase in the carrier concentration at transparency for the In$_{0.36}$Ga$_{0.64}$N$_{0.02}$As$_{0.98}$/GaAs QW laser structure and a decrease in the separation between the conduction and valence band quasi-Fermi energies, $E_{F_{\rm{c}}} - E_{F_{\rm{v}}}$, for a fixed carrier concentration. As a consequence the peak gain decreases at a fixed carrier density in the In$_{0.36}$Ga$_{0.64}$N$_{0.02}$As$_{0.98}$/GaAs structure when compared to the N-free case, Figure~\ref{fig:GaInNAs_QW_gain_n}. There is however a much weaker variation in the peak gain versus radiative current density, Figure~\ref{fig:GaInNAs_QW_gain_Jrad}. This weak variation reflects the fact that for a fixed quasi-Fermi level separation, the radiative current $J_{\rm{rad}}$ in a QW laser is approximately proportional to $m^{*}_{r} \times \vert M_{\rm{e-h}} \vert^{2}$, where $m^{*}_{r}$ is the band edge reduced effective mass. Because the VB mass, $m^{*}_{v}$, is always larger than the CB mass, $m^{*}_{c}$, the reduced mass $m^{*}_{r}$ is determined primarily by $m^{*}_{e}$. We saw above that $m^{*}_{e} \times \vert M_{\rm{e-h}} \vert^{2}$ is approximately constant, thus accounting for the calculated weak variation in $J_{\rm{rad}}$.

\begin{figure}[t!]
  \centering
  \includegraphics[scale=0.4]{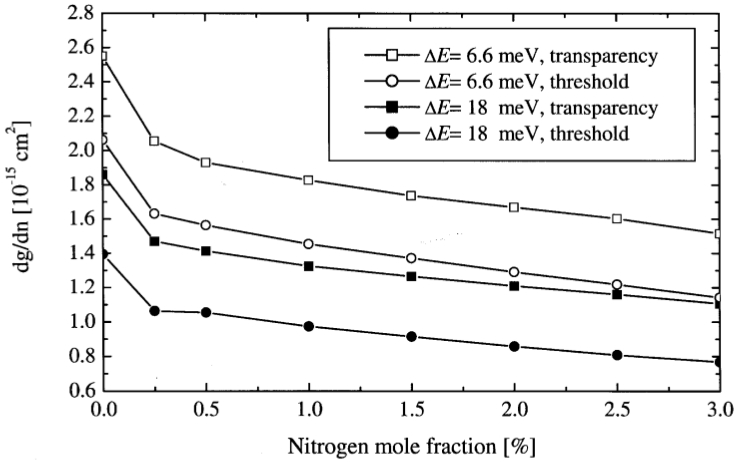}
  \caption{Calculated variation of differential gain at transparency and at threshold in a series of ideal 7 nm In$_{y}$Ga$_{1-y}$N$_{x}$As$_{1-x}$ single quantum well structures, in which the In composition $y$ was varied with N composition $x$ in order to maintain the ground state transition energy $e1$-$hh1$ constant at 0.954 eV (1.3 $\mu$m). (Reproduced from Ref.~\cite{Tomic_Photon_Tech_Lett_2003}.)}
  \label{fig:GaInNAs_differential_gain}
\end{figure}

The differential gain plays a key role in determining both the threshold carrier density and the bandwidth of a directly modulated semiconductor laser. The modulation response resonance frequency $\omega_{\rm{r}}$ is proportional to the square root of the differential gain with respect to the carrier density, $\omega_{\rm{r}} \propto \sqrt{ \rmd g / \rmd n }$. Figure~\ref{fig:GaInNAs_differential_gain} shows the calculated variation of differential gain at transparency and at threshold in a series of ideal 7 nm In$_{y}$Ga$_{1-y}$N$_{x}$As$_{1-x}$ QW structures. In order to maintain the QW ground-state transition energy $e1$-$hh1$ constant at 0.954 eV (1.3 $\mu$m), in the range of nitrogen composition from $x =$ 0\% to 3\%, we have calculated a decrease of In composition $y$ from 55\% to 30\% \cite{Tomic_Photon_Tech_Lett_2003}. We find that for a fixed broadening factor the calculated differential gain decreases with increasing N content, $x$, with the most rapid decrease observed at low $x$. As expected, the differential gain also decreases with increasing broadening factor. Because the measured broadening in GaInNAs increases with $x$ \cite{Klar_SST_2002}, we conclude that the calculated threshold $\rmd g / \rmd n$ value of $\sim$ 0.86 $\times$ 10$^{-15}$ cm$^{-1}$ for $x =$ 0.02 is only about 40\% of the value which could be achieved in an ideal GaInAs 1.3 $\mu$m QW laser. The initial rapid decrease in differential gain (already notable at the first data point, $x =$ 0.25\%) is due to the strong coupling between the N resonant states and the conduction band edge states, even for small values of $x$.

Figure~\ref{fig:GaInNAs_differential_gain} suggests that the optimal GaInNAs/GaAs QW laser device should contain minimal nitrogen, ideally being N-free! This however is not possible because of the excessively large strain required to achieve 1.3-$\mu$m emission in a GaInAs QW. We estimated, using the strain-thickness criteria in Ref.~\cite{Reilly_SST_1989}, that one needs $x >$ 1.5\%  and $y <$ 39\% to achieve 1.3-$\mu$m emission in a pseudomorphic In$_{y}$Ga$_{1-y}$N$_{x}$As$_{1-x}$/GaAs QW structure. The N content can be further reduced by adding more In to narrower QWs and also by growing tensile-strained layers above and below the QW, as demonstrated by Tansu et al., who achieved 1.3 $\mu$m emission with $x =$ 0.5\%, a quantum well width of 6 nm and utilising strain-compensating GaAs$_{0.85}$P$_{0.15}$ tensile layers in the barrier region \cite{Tansu_APL_2002,Tansu_Photon_Tech_Lett_2002}.

% Figure: GaInNAs differential gain

Comparing our theoretical results on GaInNAs/GaAs with previous theoretical analysis of 1.3 $\mu$m InGaAsP/InP and InGaAlAs/InP structures \cite{Tomic_JSTQE_2003,Seki_JQE_1994,Pan_JQE_1996}, we find that the GaInNAs material has a higher differential gain of $\sim$ 0.8 $\times$ 10$^{-15}$ cm$^{-1}$ compared to $\sim$ 0.6 $\times$ 10$^{-15}$ cm$^{-1}$ in InP-based devices for the same amount of compressive strain ($\sim$ 1.5\%), optical gain  ($\sim$ 1300 cm$^{-1}$), or well thickness ($\sim$ 7 nm), despite the significantly larger line broadening of 18 meV assumed in GaInNAs compared to 6.6 meV in the InP-based devices. Stronger optical confinement is possible in 1.3 $\mu$m GaInNAs/GaAs-based lasers compared to InP-based devices, because of the larger refractive index step achieveable through use of AlGaAs cladding layers. This reduces the number of quantum wells required in an edge-emitting device, while also opening the possibility of GaInNAs-based 1.3 and 1.5 $\mu$m VCSELs. The theoretical analysis presented here therefore confirms the potential of GaInNAs-based lasers both for edge- and surface-emitting laser applications in the telecommunication wavelength range.

%%%%%%%%%%%%%%%%%%%%%%%%%%%%%%%%%%%%%%%%%%%%%%%%%%%%%%%%%%%%%%%
%% Influence of N on loss mechanisms in ideal GaInNAs lasers %%
%%%%%%%%%%%%%%%%%%%%%%%%%%%%%%%%%%%%%%%%%%%%%%%%%%%%%%%%%%%%%%%

\subsection{Influence of N on loss mechanisms in GaInNAs lasers}
\label{sec:Influence_of_N_on_losses}

While there has been considerable interest in the influence of nitrogen incorporation on the electronic structure and optical transitions in GaInNAs/GaAs QWs, there have been comparatively few quantitative investigations of the primary loss mechanisms in lasers based on these material systems. Such studies are of
interest since they allow a direct comparison between the relative importance of radiative and non-radiative recombination paths in dilute nitride lasers compared to their conventional InP-based counterparts. Non-radiative recombination processes make a significant contribution to the threshold current in conventional InP-based lasers. This leads to a strong temperature dependence of the threshold current in such devices \cite{Phillips_JSTQE_1999}. A number of factors can contribute to this temperature dependence, such as carrier leakage due to weak electron confinement \cite{Chen_APL_1983,Barrau_JAP_1992} or inter-valence band absorption (IVBA) \cite{Adams_JJAP_1980,Seki_JQE_1996}, but the dominant loss mechanism in InP-based lasers at threshold is generally regarded to be Auger recombination \cite{Phillips_JSTQE_1999,Dutta_APL_1981}.

We begin our discussion of GaInNAs lasers with a description of the quantitative experimental analysis of recombination pathways in 1.3 $\mu$m GaInNAs/GaAs lasers carried out in Ref.~\cite{Fehse_JSTQE_2002}, highlighting the importance of defect-related and Auger recombination in such devices. We then provide an interpretation of the observed device characteristics in terms of detailed theoretical calculations of the effects of N on Auger recombination rates in GaInNAs/GaAs systems \cite{Andreev_APL_2004}. This is then used as a basis to understand the observed threshold characteristics of GaInNAs lasers emitting across the full wavelength range from 1.2 $\mu$m to 1.6 $\mu$m.

Under the assumptions of charge neutrality in the active region (electron density $n$ = hole density $p$) and negligible carrier leakage, the current density in a single quantum well (SQW) laser can be written in the Boltzmann approximation as

\begin{equation}
  \label{eq:laser_current}
  J = J_{\rm{mono}} + J_{\rm{rad}} + J_{\rm{Auger}} = e \left( An + Bn^{2} + Cn^{3} \right)
\end{equation}
where $J_{\rm{mono}}$, $J_{\rm{rad}}$ and $J_{\rm{Auger}}$ refer respectively to the current densities due to monomolecular/defect-related ($\propto n$), radiative ($\propto n^{2}$) and Auger ($\propto n^{3}$) recombination, with $A$, $B$ and $C$ the monomolecular, radiative and Auger recombination coefficients. In each case the recombination rate varies in the Boltzmann approximation as carrier density $n$ to the power of the number of carriers involved -- e.g. because radiative recombination is a bimolecular process, it varies as $n^{2}$. 

It is possible from Eq.~\ref{eq:laser_current} to express the total current density $J$ over a limited current range as $J \propto n^{z}$, with $1 \leq z \leq 3$ \cite{Higashi_JSTQE_1999,Phillips_JSTQE_1999} and with the value of $z$ varying depending on whether the dominant contribution to the current density is due to monomolecular ($z = 1$), radiative ($z = 2$) or Auger ($z = 3$) recombination.

It is also possible to measure the integrated spontaneous emission (SE) rate $L$ from a semiconductor laser by etching a window in the laser substrate and then collecting the spontaneous emission emitted through the window. Because $L$ is proportional to $Bn^{2}$, this allows the total current density $J$ to be related to the integrated spontaneous emission $L$ as $J \propto L^{\frac{z}{2}}$, which can be rewritten as $\ln J \propto z \ln L^{\frac{1}{2}}$. This then enables $z$ to be quantified experimentally for a given laser by plotting $\ln J$ against $\ln L^{\frac{1}{2}}$, with the slope of the log-log plot giving the value for $z$.

Such an analysis was applied in Ref.~\cite{Fehse_JSTQE_2002} to the study of two 1.3 $\mu$m GaInNAs/GaAs laser structures, the first containing a single In$_{0.36}$Ga$_{0.64}$N$_{0.017}$As$_{0.983}$ QW with GaAs barriers (SQW) and the second containing three of the same QWs but with In$_{0.05}$Ga$_{0.95}$N$_{0.018}$As$_{0.982}$ barriers (3QW), approximately lattice-matched to GaAs.

The temperature dependence of laser threshold current density $J_{\rm{th}}$ is often expressed in terms of the $T_{0}$ value, where $T_{0}$ is defined by $T_{0}^{-1} = \frac{\rm{d}}{\rm{d}T} \ln J_{\rm{th}}$. A low value of $T_{0}$ then implies a strong temperature dependence of the threshold current. Measurement of $L$ as a function of the injected current below threshold in the temperature range from $T =$ 290 -- 335 K in the SQW device yielded a value of $T_{0} \sim$ 310 K for the radiative contribution to the total threshold curent, which was considerably larger than the value of $T_{0} \sim$ 90 K measured for the total threshold current. This immediately reveals two important device characteristics: firstly, that the laser is dominated, below threshold by a non-radiative recombination mechanism and secondly, that this non-radiative recombination path is substantially more temperature sensitive than the radiative contribution to the threshold current.

By fitting the slope of $\ln J$ versus $\ln L^{\frac{1}{2}}$ at temperatures below 220 K, a slope value was determined for the SQW laser at threshold of $z_{\rm{th}} = 1.6$. This lies between the values for purely radiative and for purely defect-related recombination. When compared to the measured value of $z_{\rm{th}} =$ 2.1 over the same temperature range in a 1.3 $\mu$m InGaAsP/InP device \cite{Phillips_JSTQE_1999} it shows that whereas defect-related recombination is not an important loss mechanism in conventional InP-based devices, it can play a major role in GaInNAs/GaAs lasers.

% Figure: Variation of threshold current density as a function of temperature for the SQW GaInNAs laser

\begin{figure}
    \centering
    \includegraphics[scale=0.35]{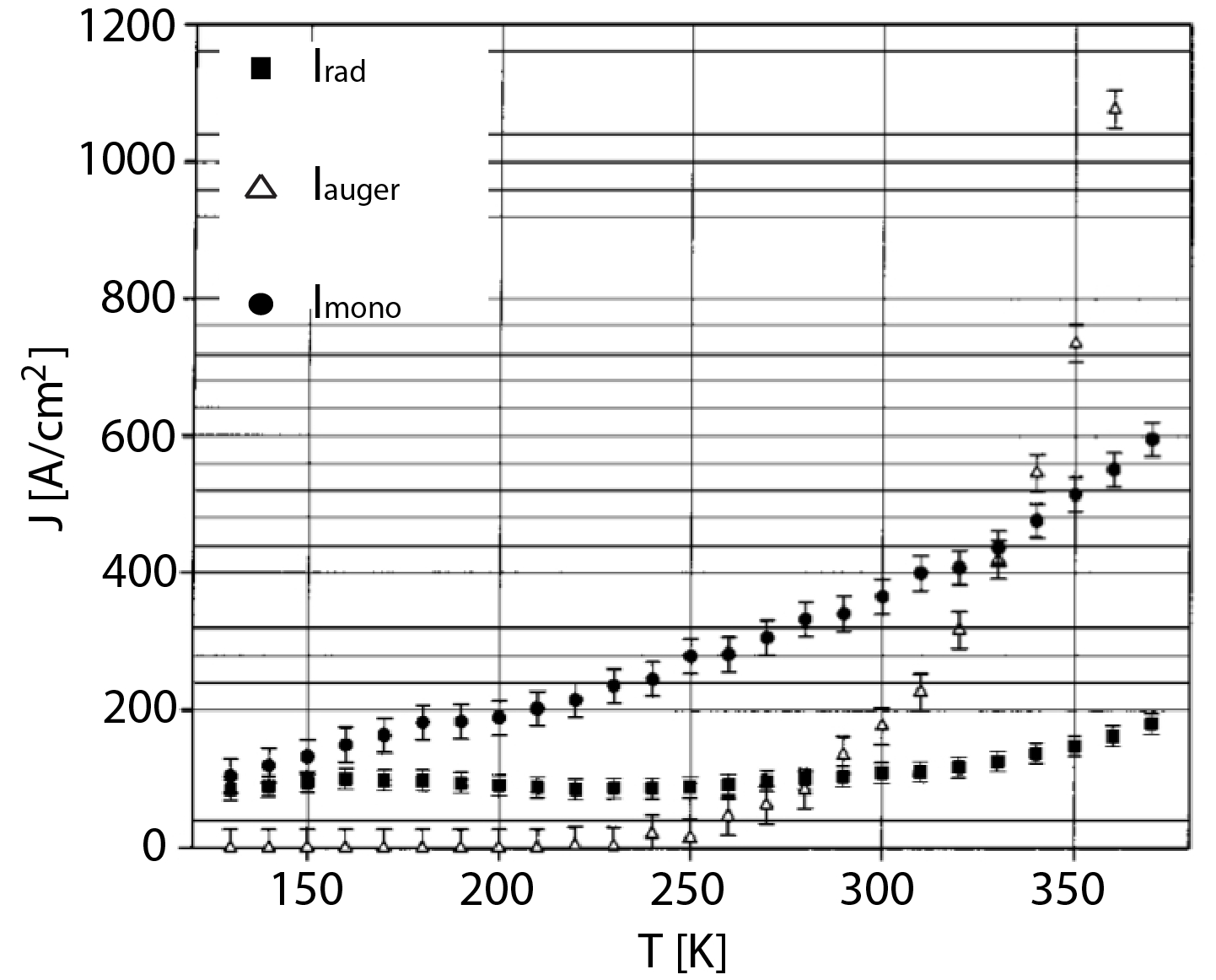}
    \caption{Experimentally measured variation of the monomolecular, radiative and Auger current densities at threshold, as a function of temperature for a Ga$_{0.64}$In$_{0.36}$N$_{0.017}$As$_{0.983}$ SQW laser device, as determined from the analysis outlined in the text. (Reproduced from Ref.~\cite{Fehse_JSTQE_2002}.)}
    \label{fig:GaInNAs_QW_J_vs_T_all_paths}
\end{figure}

By careful analysis of the data measured over a wider temperature range, it was shown that at low temperatures ($T \lesssim$ 200 K) defect-related recombination constituted $\sim$ 70\% of the total threshold current in the device considered and contributed $\sim$ 55\% of the total current at room temperature. A full analysis indicated that in the SQW laser defect-related, radiative and Auger currents constitute respectively $\sim$ 55\%, 20\% and 25\% of $J_{\rm{th}}$ at room temperature.
The measured variation of each of the monomolecular, radiative and Auger current paths in the SQW device are shown in Figure~\ref{fig:GaInNAs_QW_J_vs_T_all_paths}. It can be seen that the Auger contribution to the threshold current increases rapidly at higher temperatures, becoming the dominant current path in the given device above $T=$ 350K. This is confirmed by the measured value of $z_{\rm{th}}$ = 2.8 at $T = $ 370 K. Comparing this to the value of $z_{\rm{th}}$ = 2.9 for the 1.3 $\mu$m InGaAsP/InP device of Ref.~\cite{Phillips_JSTQE_1999} shows that the GaInNAs/GaAs material system does not differ significantly in its Auger-dominated high temperature characteristics from conventional InP-based lasers.

A broadly similar behaviour was obtained in the 3QW device \cite{Fehse_JSTQE_2002}. Increasing the number of QWs to three reduces the modal gain required per well (assuming a fixed loss level) and therefore should also reduce the threshold carrier density per well, $n_{\rm{th}}$. This then reduces the relative importance of $J_{\rm{Auger}}$ compared to $J_{\rm{rad}}$:  at the lower $n$ value, $Bn^{2}$ is larger compared to $Cn^{3}$ than it is at the higher $n_{\rm{th}}$ value. However, the reduced value of $n_{\rm{th}}$  simultaneously increases the relative importance of the defect-related recombination, since $An$ becomes larger compared to $Bn^{2}$ than at larger carrier densities. The measured value of $z_{\rm{th}} =$ 2.4 at $T =$ 370 K in the 3QW laser is then consistent with the increased relative importance of defect-related compared to Auger recombination in this device.

% Figure: (a) Schematic illustrations of Auger recombination processes

\begin{figure}
    \centering
    \includegraphics[scale=0.50]{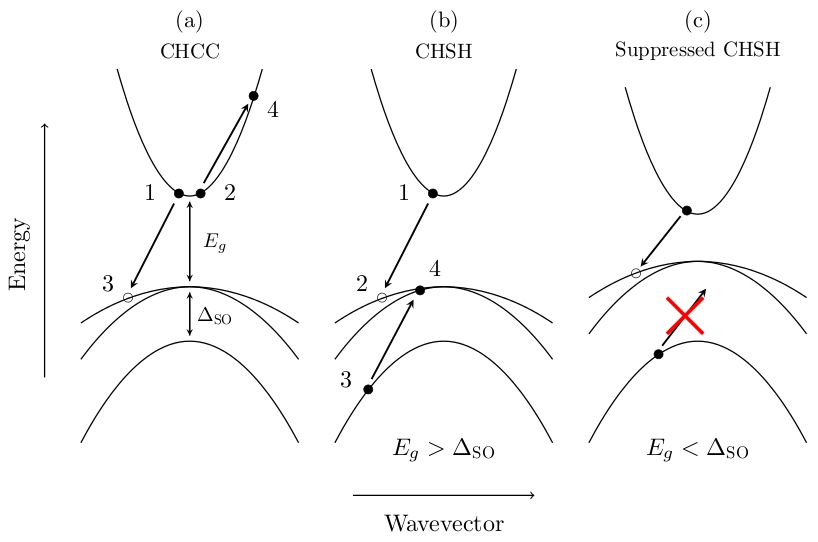}
    \caption{(a) Schematic illustration of a CHCC Auger recombination process in which a conduction electron (1) recombines with a valence hole (3), with the released energy exciting a second conduction electron (2) to a vacant higher conduction state (4), such that energy and momentum are conserved. (b) Schematic illustration of a CHSH Auger recombination process in which a conduction electron (1) recombines with a valence hole (2), with the released energy exciting a valence hole (4) to the spin-split-off band (3), such that energy and momentum are conserved. (c) Schematic illustration of (b) in the case where the spin-orbit splitting exceeds the band gap, $\Delta_{\scalebox{0.7}{\rm{SO}}} > E_{g}$. For $\Delta_{\scalebox{0.7}{\rm{SO}}} > E_{g}$ the CHSH process is forbidden by conservation of energy, since the energy required to excite a hole to the spin-split-off band exceeds the energy produced by the electron-hole recombination. }
    \label{fig:Auger_schematic}
\end{figure}

The relative importance of Auger and of radiative recombination in these SQW and 3QW lasers can be explained theoretically on the basis of models employing a 10-band \textbf{k}$\cdot$\textbf{p} Hamiltonian for GaInNAs \cite{Tomic_JSTQE_2003,Fehse_JSTQE_2002}. Two main Auger recombination processes are considered in telecomm lasers, namely the CHSH process discribed earlier and the CHCC process, where a \textbf{C}onduction electron recombines with a \textbf{H}eavy hole, exciting an electron from near the bottom of the \textbf{C}onduction band to a higher energy \textbf{C}onduction band state. Both the CHCC and CHSH processes are depicted schematically in Figure~\ref{fig:Auger_schematic}. Previous experimental and theoretical analysis has shown that the CHSH process is dominant in conventional InP-based lasers, providing a current path which typically accounts for the majority of the Auger current \cite{Higashi_JSTQE_1999,Sweeney_PSSb_1999,Silver_JQE_1998}. 

A detailed theoretical study of the effects of N on a GaInNAs/GaAs QW structure \cite{Andreev_APL_2004} confirmed that the contribution of Auger recombination is relatively unchanged in a GaInNAs laser compared to that observed in conventional InP-based lasers. Because nitrogen incorporation in GaInAs only perturbs the VB structure weakly, the CHSH Auger recombination mechanism, involving the excitation of a hole from the heavy-hole to the spin-split off band, again dominates in the GaInNAs material systems, with the CHSH recombination rate in GaInNAs found to be comparable to that in the equivalent N-free system \cite{Andreev_APL_2004}.

Based on this experimental \cite{Fehse_JSTQE_2002} and theoretical \cite{Andreev_APL_2004} evidence it is clear that Auger recombination is as important in 1.3 $\mu$m GaInNAs lasers as in their InP-based counterparts.

By increasing the N content, it is possible to extend the emission of GaInNAs-based lasers to longer wavelengths. The first emission beyond 1.5 $\mu$m with GaInNAs/GaAs based devices was reported in July 2000 by Fischer et al. using a 5\% N and 38\% In double quantum well structure \cite{Fischer_EL_2000}. However, the threshold current density of these devices was very large, with $J_{\rm{th}} \sim $ 60 kA cm$^{-2}$ at $\lambda =$ 1.51 $\mu$m. The development of GaInNAs 1.5 $\mu$m devices and the reduction in the reported threshold current was then rapid, with further improvements achieved by introducing Sb to give GaInNAsSb QW systems. The development of the GaInNAs material system was largely driven by Infineon. In 2004 they reported long wavelength devices at $\lambda=1.4$ $\mu$m with $J_{\rm{th}} =$ 690 A cm$^{-2}$ and at $\lambda =$ 1.43 $\mu$m with $J_{\rm{th}} =$ 1090 A cm$^{-2}$ \cite{Averbeck_ISLC_2004}. At the time these values were the lowest values for Sb-free devices emitting beyond 1.3 $\mu$m. By 2005 they had extended the emission to 1.51 $\mu$m with a GaInNAs SQW and GaAs barriers \cite{Averbeck_JCG_2005}. This device had a record low threshold current density for any GaAs-based device at this wavelength with $J_{\rm{th}} =$ 780 A cm$^{-2}$.

The development of the GaInNAsSb system was largely driven by the Harris group at Stanford who extended the emission wavelength to 1.55 $\mu$m in 2005 \cite{Bank_APL_2005}. With further developments of their device design and optimisation of their growth the Stanford group produced devices with a record low threshold current density at 1.55 $\mu$m of 579 A cm$^{-2}$ \cite{Bank_EL_2006} using GaInNAsSb and GaNAs barriers in 2006. These devices remain the benchmark for 1.55 $\mu$m emission using the GaInNAs(Sb) material system. The Harris group led the use of antimony in the growth of GaInNAs active regions. The general consensus is that antimony acts as a reactive surfactant enhancing the incorporation of N \cite{Massies_PRB_1993}. In addition, it allows higher growth temperatures, enabling improved optical quality, because of the formation of fewer point defects \cite{Bank_PhD_2006}.

\begin{figure}[t!]
  \centering
  \hspace{-1.0cm}
  \subfigure[]{ \vspace{1.0cm} \includegraphics[scale=0.21]{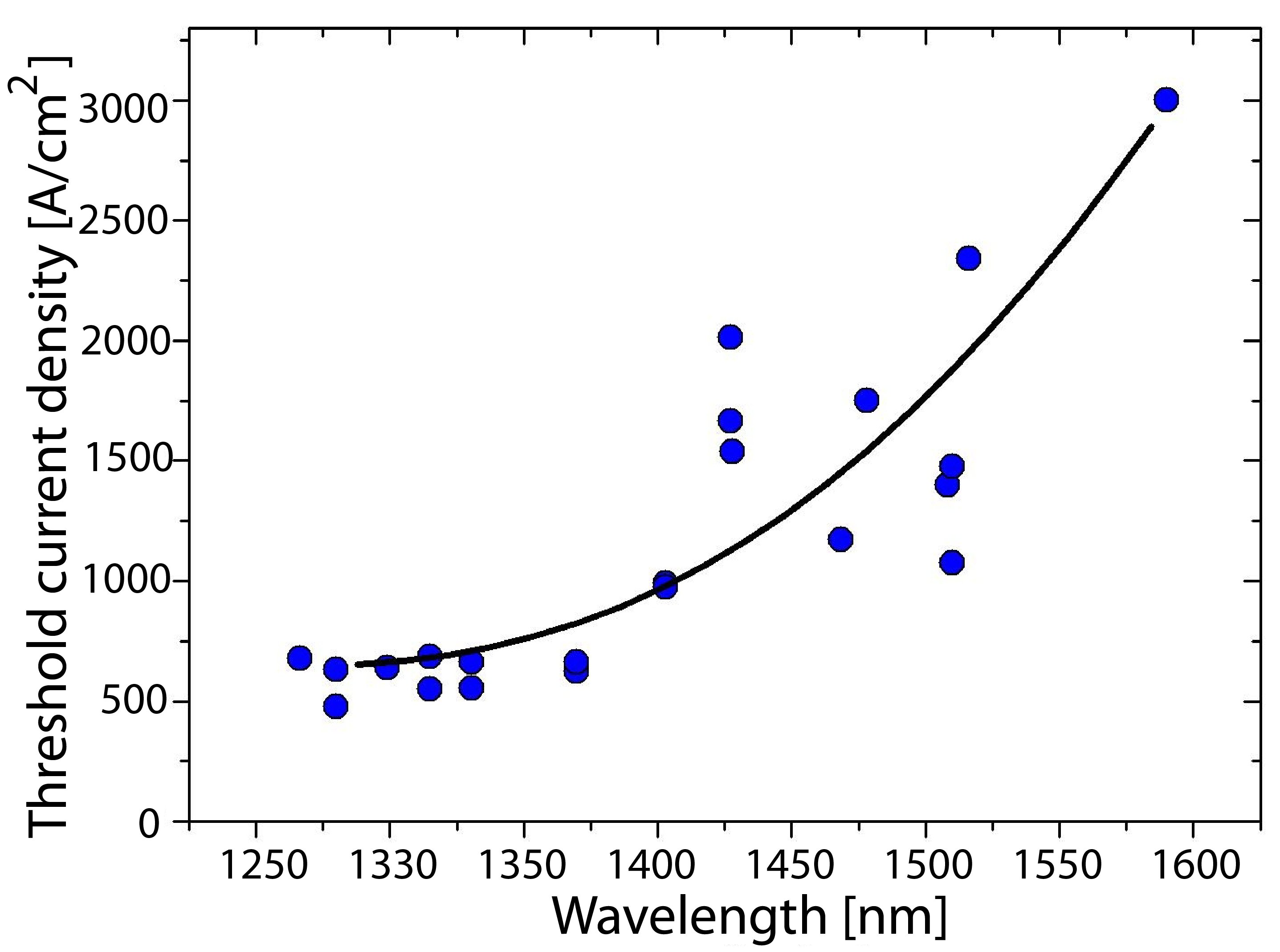} \label{fig:Jth_vs_wavelength_GaInNAs_room_temp} }
  \hspace{-0.5cm}
  \subfigure[]{ \includegraphics[scale=0.21]{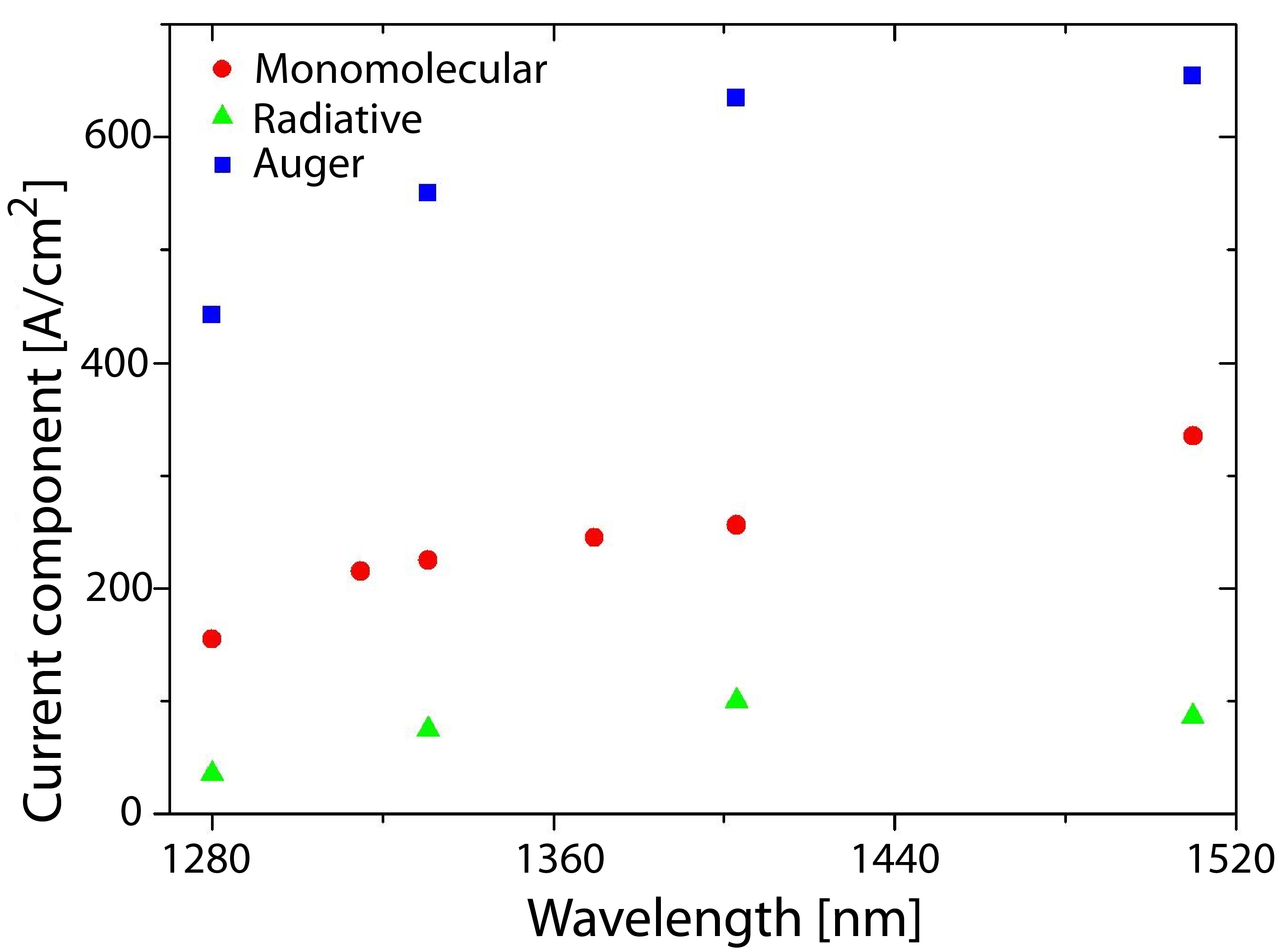} \label{fig:GaInNAs_current_recombination_paths} }
  \caption{(a) Variation of threshold current density as a function of wavelength in a series of GaInNAs devices with increasing N fraction, (b) corresponding variation of the individual current paths as a function of wavelength. The increase in threshold current with wavelength is attributable to an increase both in defect-related and in Auger recombination.}
\end{figure}

McConville et al. \cite{McConville_PhD_2007} in collaboration with Infineon undertook a study of threshold current density as a function of wavelength for GaInNAs/GaAs based devices. The emission wavelength of the devices spanned from 1.27 $\mu$m to 1.6 $\mu$m, with this range achieved by varying the nitrogen and indium contents within the quantum well from 1.3\% to 4.5\% and from 30\% to 40\% respectively. Figure~\ref{fig:Jth_vs_wavelength_GaInNAs_room_temp} shows the corresponding variation of $J_{\rm{th}}$ at room temperature as a function of wavelength. It is clear that the threshold current density increases with increasing wavelength. Using a spontaneous emission analysis similar to that described earlier, McConville quantified the contribution of each of the current paths to the threshold current density in a series of devices, as shown in Figure~\ref{fig:GaInNAs_current_recombination_paths}. It is clear from Figure~\ref{fig:GaInNAs_current_recombination_paths} that the increase in threshold current is due to an increase in both the monomolecular and the Auger currents with increasing wavelength, with little change observed in the radiative current measured in the different devices. Auger recombination evidently persists over the entire wavelength range investigated.

% Figure: (a) Threshold current density as a function of wavelength in the Infineon GaInNAs lasers, (b) Breakdown of recombination paths in the 1280 - 1510 nm range

Auger recombination places intrinsic limits on the threshold characteristics and temperature stability of GaInNAs lasers (and conventional InP-based devices) operating at telecomm wavelengths. This then raises the challenge as to whether it could be possible to eliminate Auger losses and temperature sensitivity in lasers operating in the 1.3 and 1.5 $\mu$m wavelength ranges. To address this question, we now turn our attention to an emerging class of highly-mismatched III-V alloys, namely dilute bismide alloys and the promise they offer in terms of band structure engineering for highly efficient and thermally stable optoelectronic properties.

%%%%%%%%%%%%%%%%%%%%%%%%%%%%%%%%%%%%%%%%%%%%%%%%%%%%%%%%%%%%%%%%%%%%%%%%%%%%%%%%%%%%%%%%
%% Dilute bismide alloys for highly efficient, temperature stable photonic components %%
%%%%%%%%%%%%%%%%%%%%%%%%%%%%%%%%%%%%%%%%%%%%%%%%%%%%%%%%%%%%%%%%%%%%%%%%%%%%%%%%%%%%%%%%

\section{Dilute bismide alloys for highly efficient, temperature stable photonic components}
\label{sec:dilute_bismides}

We have described that telecomm lasers based both on dilute nitride and on conventional III-V alloys suffer from significant intrinsic losses, particularly above room temperature. The observed degradation in device performance with increasing temperature is primarily due to the presence of Auger recombination and of IVBA processes \cite{Sweeney_PSSb_1999}. By investigation of antimony-containing alloys such as InGaAsSb/GaSb it has been demonstrated that antimonide-based laser devices offer benefits such as reduced threshold current and temperature sensitivity in the 2 -- 3 $\mu$m wavelength range \cite{Sweeney_ICTON_2011}. These improved device characteristics have been attributed to the spin-orbit-splitting, $\Delta_{\scalebox{0.7}{\rm{SO}}}$, exceeding the band gap, $E_{g}$. This leads to a suppression of the dominant CHSH Auger recombination mechanism due to the inability of such transitions to conserve energy for $E_{g} < \Delta_{\scalebox{0.7}{\rm{SO}}}$ (cf. Figure~\ref{fig:Auger_schematic}), and should also eliminate intervalence band absorption \cite{Sweeney_ICTON_2011,O'Brien_APL_2006}. Similar benefits have also been observed in GaInAsSbP mid-infrared light emitting diodes \cite{Cheetham_APL_2011}.

Achieving similar characteristics at telecomm wavelengths has remained an elusive task - it has not been possible to effectively suppress Auger recombination at high temperatures. Sweeney et al. recently proposed that by engineering the band structure to achieve $ \Delta_{\scalebox{0.7}{\rm{SO}}} > E_{g}$, it should be possible to realise laser devices operating at telecommunication wavelengths with suppressed Auger recombination and IVBA and hence improved efficiency and temperature stablity \cite{Sweeney_ICTON_2011,Sweeney_patent_2010}.

% Figure: (a) Band edge energies and (b) energy gaps in GaBiAs

\begin{figure}[t!]
  \centering
  \includegraphics[scale=0.4]{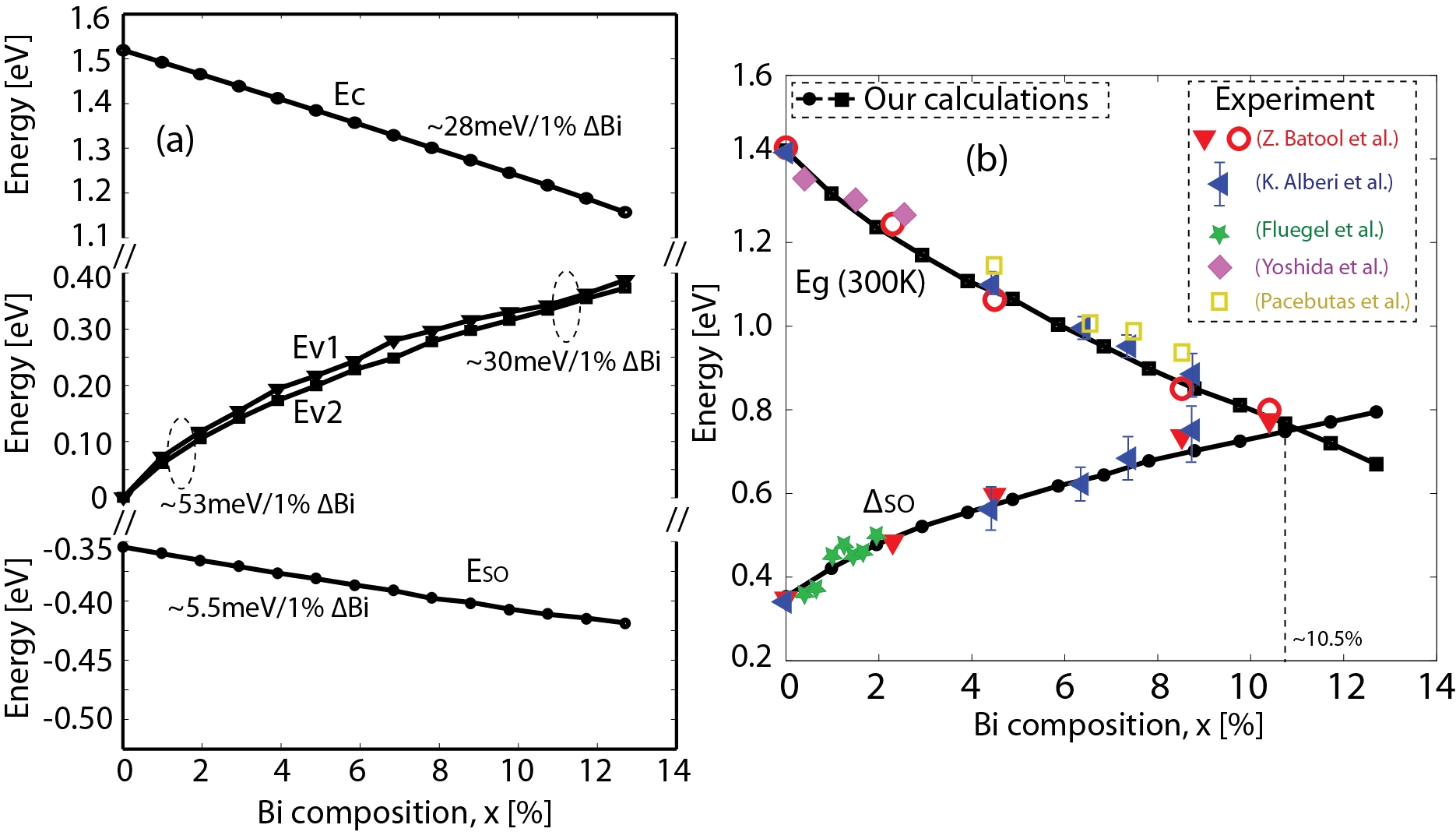}
  \caption{(a) Calculated variation of the band edge energies of the lowest conduction band, top two valence bands and spin-split-off band, with Bi composition in GaBi$_{x}$As$_{1-x}$ using an sp$^{3}$s$^{*}$ tight-binding model. (b) Solid lines/Closed black points: Calculated variation the band gap ($E_{g}$) and spin-orbit-splitting ($\Delta_{\scalebox{0.7}{\rm{SO}}}$), as a function of Bi composition in GaBi$_{x}$As$_{1-x}$ using an sp$^{3}$s$^{*}$ tight-binding model. Various symbols: Experimental measurements of $E_{g}$ and $\Delta_{\scalebox{0.7}{\rm{SO}}}$ from the indicated sources. (Reproduced from Ref.~\cite{Usman_PRB_2011}.)}
  \label{fig:GaBiAs_band_edges}
\end{figure}

Such a band structure could be achieved using alloys containing dilute amounts of bismuth (Bi), the largest stable group V element. Bismuth has a very large spin-orbit splitting of $\approx$ 2.2 eV and it has been demonstrated that incorporating dilute quantities of Bi in GaAs to form GaBi$_{x}$As$_{1-x}$ can give some highly interesting and unusual electronic properties. 

Experimental investigations of GaBi$_{x}$As$_{1-x}$ alloys have revealed (i) a large downward bowing of the band gap, which reduces by $\approx$ 90 meV per percent Bi for $x \lesssim 5$\% \cite{Sweeney_ICTON_2011,Usman_PRB_2011,Tixier_APL_2005,Alberi_PRB_2007} and (ii) a large upward bowing of the spin-orbit-splitting. It has been demonstrated that these effects lead to the onset of a $ \Delta_{\scalebox{0.7}{\rm{SO}}}> E_{g}$ regime in the alloy for $x \approx 10$\% \cite{Sweeney_ICTON_2011,Usman_PRB_2011}. Based on a comparison with the antimonides this latter characteristic is of great technological interest, opening up an avenue to suppress CHSH Auger recombination processes at sufficiently high Bi compositions. This could enable highly efficient photonic devices operating with reduced threshold current and temperature sensitivity at telecommunication wavelengths.

% Figure: Composition space charts for GaBiNAs/GaAs

\begin{figure}[t!]
  \centering
  \subfigure[]{ \includegraphics[scale=0.92]{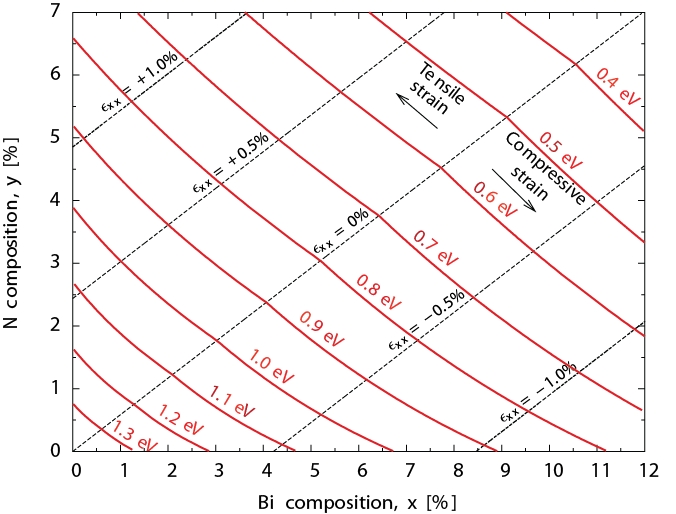} \label{fig:GaBiNAs_band_gap} }
  \subfigure[]{ \includegraphics[scale=0.92]{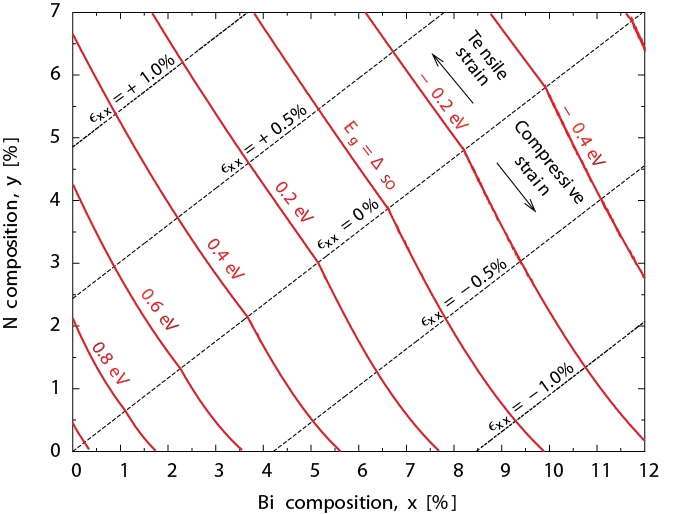} \label{fig:GaInNAs_spin_orbit_splitting_minus_band_gap} }
  \caption{Calculated room temperature variation of (a) the band gap and (b) the difference between the band gap and spin-orbit-splitting ($E_{g} - \Delta_{\scalebox{0.7}{\rm{SO}}}$), as a function of Bi and N composition, $x$ and $y$, for GaBi$_{x}$N$_{y}$As$_{1-x-y}$ epitaxially grown on GaAs. The solid lines in (a) denote paths in the composition space along which the band gap is constant; dashed lines denote paths along which the strain is constant. The solid lines in (b) denote paths in the composition space along which $E_{g} - \Delta_{\scalebox{0.7}{\rm{SO}}}$ is constant; dashed lines denote paths along which the strain is constant. Regions in (b) where $\Delta_{\scalebox{0.7}{\rm{SO}}} > E_{g}$ (to the right of the $E_{g} = \Delta_{\scalebox{0.7}{\rm{SO}}}$ contour) indicate alloys in which suppression of CHSH Auger recombination and intervalence band absorption can be expected to occur.}
\end{figure}

Bismuth, being the largest stable group V element, is significantly larger and more electropositive than As. It should therefore be expected Bi could give rise to impurity levels, as was the case for N. However, any Bi-related impurity states in GaAs should lie near or above the valence band edge (VBE) and, if an anti-crossing interaction occurs, it will occur between the Bi-related impurity levels and the VBE of the host (GaAs) matrix. It has been proposed that the band structure of GaBi$_{x}$As$_{1-x}$ can be explained in terms of a valence band BAC interaction \cite{Usman_PRB_2011,Alberi_PRB_2007,Alberi_APL_2007,Broderick_12band_2012}. The presence of Bi-related states has been confirmed by several theoretical studies \cite{Usman_PRB_2011,Zhang_PRB_2005}, but there remains controversy as to whether or not the observed bowing of the band gap and spin-orbit-splitting with increasing Bi composition can be attributed to a BAC interaction occuring in the valence band in GaBi$_{x}$As$_{1-x}$ \cite{Deng_PRB_2010}. Strong evidence for the BAC model in GaNAs is provided by the observation in photoreflectance of the $E_{+}$ transitions shown in Figure~\ref{fig:GaNAs_PR}. No equivalent transitions have been found in GaBiAs.

We presented in Ref.~\cite{Usman_PRB_2011} a detailed investigation of the electronic structure of GaBi$_{x}$As$_{1-x}$ based on an sp$^{3}$s$^{*}$ tight-binding model. The model reproduces the measured variation of $E_{g}$ and $\Delta_{\scalebox{0.7}{\rm{SO}}}$ with Bi composition in GaBi$_{x}$As$_{1-x}$ across the experimentally investigated composition range, to a high degree of accuracy and the observed crossover to a $ \Delta_{\scalebox{0.7}{\rm{SO}}} > E_{g}$ regime in the alloy for $x \approx 10$\%.

Based on this tight-binding model we have been able to gain several key insights into the electronic structure of GaBi$_{x}$As$_{1-x}$ and related alloys. Firstly, by considering GaP:Bi and explicitly demonstrating the validity of the BAC model as applied to the dilute bismide case, we showed that a BAC description of the GaBi$_{x}$As$_{1-x}$ VBE is justified \cite{Broderick_12band_2012}. Secondly, we demonstrated that the absence of Bi-related features in spectroscopic measurements of GaBi$_{x}$As$_{1-x}$ is attributable to a strong broadening of the Bi-related states by the large density of GaAs valence states with which they are resonant (see Figure 9 in Ref.~\cite{Usman_PRB_2011}). Thirdly, detailed calculations on large ordered and disordered supercells showed that the large bowing of $E_{g}$ with Bi composition arises not only from a strong upward shift of the alloy VBE with composition (due to anticrossing with lower lying Bi-related states), but that a significant conventional alloy reduction in the CBE energy also contributes to the observed reduction in the band gap. Finally, the spin-split-off band edge was found to admit a description in terms of a conventional alloy model, similar to the CBE, but varying less strongly in energy with Bi composition. The observed strong bowing of $\Delta_{\scalebox{0.7}{\rm{SO}}}$ is then attributed primarily to the BAC-induced upward shift of the VBE \cite{Usman_PRB_2011,Broderick_12band_2012}. Figure~\ref{fig:GaBiAs_band_edges}(a) shows the predicted variation in the band edge energies calculated using the tight-binding model. The resulting variation in the band gap and spin-orbit-splitting in a series of large, disordered, free-standing GaBi$_{x}$As$_{1-x}$ supercells is compared with experimental data in Figure~\ref{fig:GaBiAs_band_edges}(b).
 
Based on the tight-binding results, we have extended the 8-band \textbf{k}$\cdot$\textbf{p} model to include four Bi-related resonant states in a 12-band \textbf{k}$\cdot$\textbf{p} model for (In)GaBi$_{x}$As$_{1-x}$. Calculations based on this \textbf{k}$\cdot$\textbf{p} model \cite{Broderick_12band_2012} provide results in excellent agreement with PR measurements \cite{Sweeney_ICTON_2011,Batool_GaBiAs_2012} of $E_{g}$ and $\Delta_{\scalebox{0.7}{\rm{SO}}}$ for a series of MBE-grown epitaxial GaBi$_{x}$As$_{1-x}$/GaAs layers. This modified \textbf{k}$\cdot$\textbf{p} model is currently being applied to further investigate the electronic and optical properties of dilute bismide alloys and the potential of optimised bismide-based photonic components operating at telecommunication wavelengths.

The tight-binding analysis has also been extended to the quaternary dilute bismide-nitride alloy GaBi$_{x}$N$_{y}$As$_{1-x-y}$ \cite{Broderick_GaBiNAs_2012}. Co-alloying N and Bi in GaAs offers further potential for band structure engineering to design highly efficient and temperature insensitive photonic components operating at telecomm and longer wavelengths. The large variation in the GaN$_{x}$As$_{1-x}$ and GaBi$_{x}$As$_{1-x}$ band gaps with nitrogen and bismuth composition enables the GaBi$_{x}$N$_{y}$As$_{1-x-y}$ material system to span a wide wavelength range,  even when grown on a GaAs substrate. Since N and Bi in GaAs introduce, respectively, tensile and compressive strain, the N and Bi compositions in GaBi$_{x}$N$_{y}$As$_{1-x-y}$ can be chosen to lattice-match the alloy to a GaAs substrate, while allowing access to the 1.3 $\mu$m and 1.5 $\mu$m wavelength ranges and beyond. Theoretical calculations have shown that GaBi$_{x}$N$_{y}$As$_{1-x-y}$ retains the large upward bowing of the spin-orbit-splitting present in GaBi$_{x}$As$_{1-x}$ \cite{Broderick_GaBiNAs_2012}. Because co-alloying Bi with N causes a giant reduction in the band gap, this offers the possibility to eliminate the CHSH Auger recombination mechanism in lasers operating across a wide wavelength range, from 1.5 $\mu$m through to mid-infrared wavelengths \cite{Jin_APL_2012}.

By examining in detail the effects of co-alloying N and Bi in a series of ordered GaBi$_{x}$N$_{y}$As$_{1-x-y}$ supercells using the tight-binding model, we showed that the effects of Bi and N on the GaAs electronic structure are largely decoupled from each other \cite{Broderick_12band_2012,Broderick_GaBiNAs_2012,Broderick_ICTON_2011}. This indicates that the GaBi$_{x}$N$_{y}$As$_{1-x-y}$ electronic structure admits analysis in terms of separate N- and Bi- induced BAC interactions occurring in the conduction and valence bands, respectively \cite{Broderick_12band_2012}. Based on this, we have derived a 14-band \textbf{k}$\cdot$\textbf{p} model of GaBi$_{x}$N$_{y}$As$_{1-x-y}$, which is in good agreement with tight-binding calculations on both ordered \cite{Broderick_12band_2012} and disordered \cite{Broderick_GaBiNAs_2012} GaBi$_{x}$N$_{y}$As$_{1-x-y}$ supercells. Figure~\ref{fig:GaBiNAs_band_gap} shows the variation of the band gap as a function of Bi and N composition for GaBi$_{x}$N$_{y}$As$_{1-x-y}$ epitaxially grown on GaAs, calculated using the 14-band \textbf{k}$\cdot$\textbf{p} model of Ref.~\cite{Broderick_12band_2012}. This shows the wide wavelength range accessible using GaBi$_{x}$N$_{y}$As$_{1-x-y}$ grown on a GaAs substrate. Figure~\ref{fig:GaInNAs_spin_orbit_splitting_minus_band_gap} shows the calculated difference between the spin-orbit-splitting and the band gap energy over the same composition range, showing the regions in which $ \Delta_{\scalebox{0.7}{\rm{SO}}} > E_{g}$ and suppression of CHSH Auger recombination can be expected to occur.

% Figure: InGaBiAs band gap and spin-orbit-splitting

\begin{figure}[t]
    \centering
    \includegraphics[scale=0.35]{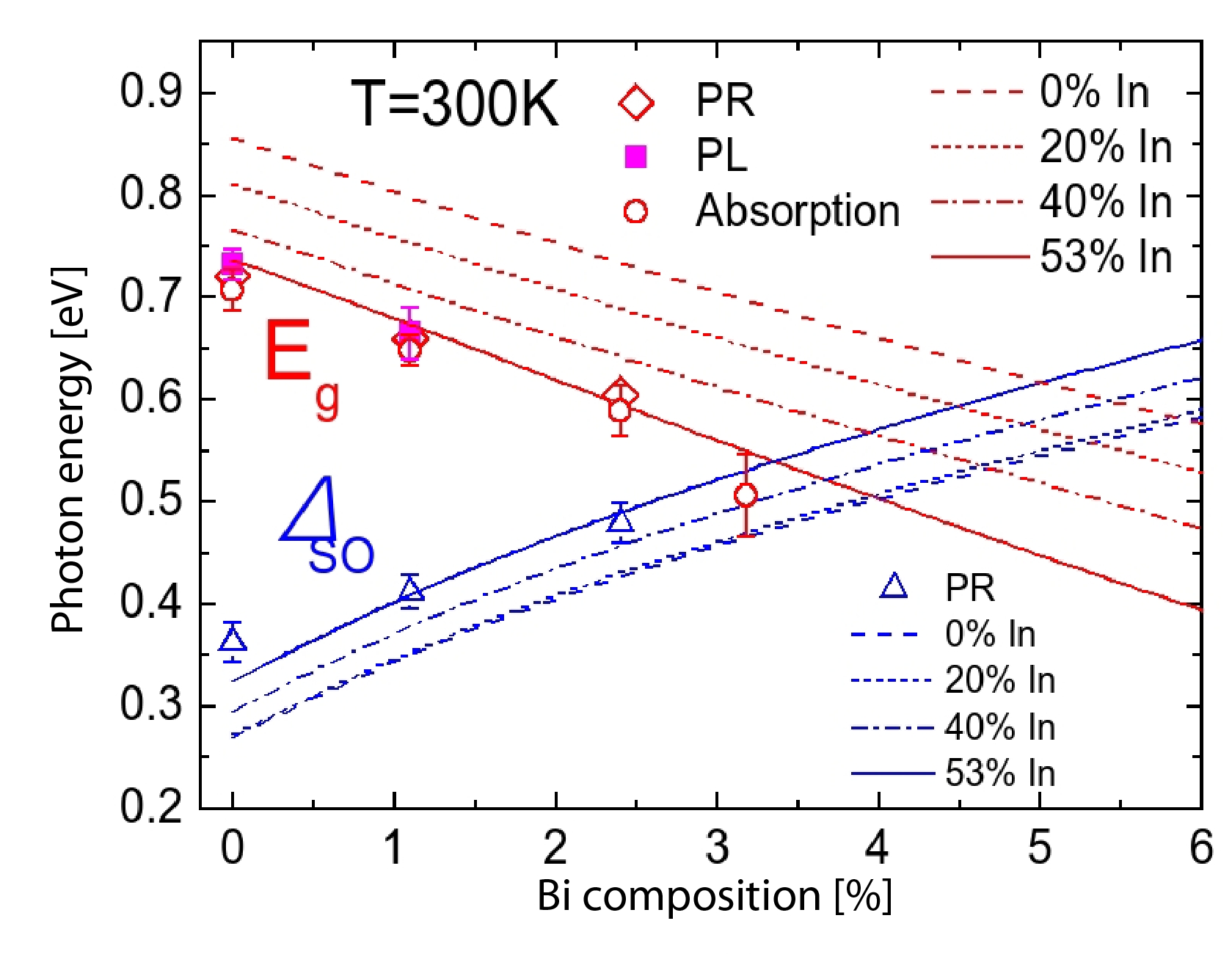}
    \caption{Measured variation in the band gap and spin-orbit splitting energies of InGaAsBi alloys grown on InP using multiple spectroscopic techniques. Also shown are the calculated variations of the band gap and spin-orbit splitting as a function of Bi and In fraction \cite{Marko_APL_2012}.}
    \label{fig:InGaBiAs_band_gap_spin_orbit_splitting}
\end{figure}

In addition to GaBi$_{x}$N$_{y}$As$_{1-x-y}$/GaAs, the In$_{1-x}$Ga$_{x}$Bi$_{y}$As$_{1-y}$/InP system offers similar promise for mid-infrared applications. Figure~\ref{fig:InGaBiAs_band_gap_spin_orbit_splitting} shows the calculated and measured variation of $E_{g}$ and $\Delta_{\scalebox{0.7}{\rm{SO}}}$ as a function of In and Bi composition for material lattice-matched to InP \cite{Marko_APL_2012}. Here it is seen that the spin-orbit-splitting exceeds the band gap for $\sim$ 3 -- 4\% bismuth. This illustrates the potential of this system to cover a wavelength range in the near-infrared, while being close to lattice matched with InP. 

The prospect of achieving $\bigtriangleup_{SO} >$ E$_{g}$ at telecomm wavelengths has sparked increasing interest in the growth, characterization, and understanding of dilute bismide alloys. High quality growth of GaBi$_x$As$_{1-x}$ on a GaAs substrate has been reported using molecular beam epitaxy (MBE) by Lu \textit{et al.} \cite{Lu_APL_2009}, showing Bi incorporation of up to $\sim$10\%. This is just sufficient to shift the band gap energy into the 1.5 $\mu$m range with $\bigtriangleup_{SO} >$ E$_{g}$, as confirmed by the photo-modulated reflectance (PR) measurements of Batool \textit{et al.} \cite{Batool_JAP_2012} included in figure 9. Room temperature photoluminescence (PL) was obtained from these samples across the full range of compositions, with the maximum PL efficiency in samples with $x$ $\sim$4.5\% consistent with reports from other growth groups. By reducing the growth temperature, further MBE samples have been grown with $x$ as high as 20\%, but with no PL signal observed from the bismide layers \cite{Tiedje_2012}. In other work, Mazur \textit{et al.} \cite{Mazur_Nano_2011} have reported the growth of an 11 nm GaBi$_x$As$_{1-x}$/GaAs ($x \sim$ 6\%) quantum well (QW) structure with low dislocation density confirmed by high-resolution x-ray diffraction, exhibiting low temperature PL with a linewidth of $\sim$40 meV. More recently, Ludewig \textit{et al}.\cite{Ludewig_JCG_2012} have reported the droplet-free epitaxial growth of GaBi$_x$As$_{1-x}$ multiple quantum well (MQW) structures on a GaAs (001) substrate by metal organic vapor phase epitaxy (MOVPE) using all-liquid group V precursors under controlled growth conditions. High resolution x-ray diffraction (HR-XRD), transmission electron microscopy (TEM) and atomic force microscopy (AFM) measurements showed that the MQW structures grown had good crystalline quality, with bismuth concentrations of up to 4.2\%. Room temperature PL was observed, where the peak position shifted to lower energies and the integrated PL signal decreased with increasing Bi fraction. Zhong \textit{et al.} \cite{Zhong_APL_2012} have demonstrated the growth of In$_x$Ga$_{1-x}$Bi$_y$As$_{1-y}$ layers on InP where the bismide layers contained up to 6.75\% Bi. In terms of progression towards devices, GaAsBi/GaAs light emitting diode structures have been fabricated with low ($<$ 2\%) Bi fraction exhibiting electroluminescence (EL) at room temperature \cite{Nadir_APL_2012}. However, the device performance was limited by defect-related recombination and carrier leakage due to the small conduction band offset. Optically pumped lasing has been demonstrated from a bulk-like GaBi$_{0.025}$As$_{0.975}$ layer sandwiched between GaAs confinement layers \cite{Tominaga_ISLC_2010}. Overall, the growth and characterization of high quality bismide samples, primarily on GaAs substrates has seen rapid progress over the last few years. There is now a strong effort by a number of groups towards the demonstration of electrically-pumped lasing in GaAsBi/GaAs-based devices and emerging interest in the growth of bismides on other substrates, such as InP, InAs and GaSb for applications into the mid-infrared.

A dedicated international meeting on bismuth-containing materials and devices has been held every year since 2010 to provide a forum for discussions about current challenges and progress in this exciting and growing area of semiconductor research~\cite{Bismides_meeting}.

%%%%%%%%%%%%%%%%%%%%%%%%%%%%%%%%
%% Discussion and Conclusions %%
%%%%%%%%%%%%%%%%%%%%%%%%%%%%%%%%

\section{Discussion and conclusions}
\label{sec:discussion_and_conclusions}

In summary, we have reviewed that highly mismatched semiconductor alloys such as GaN$_{x}$As$_{1-x}$ and GaBi$_{x}$As$_{1-x}$  have several novel electronic properties, including a rapid reduction in energy gap with increasing $x$. This has opened the possibility to achieve longer wavelength emission on GaAs. Theoretical calculations have shown that the gain and loss characteristics of ideal dilute nitride lasers should be at least as good as those of conventional InP-based telecomm lasers. Substantial progress was made in the development of GaInNAs telecomm lasers, leading to the demonstration of 1.3 $\mu$m edge-emitting and vertical cavity lasers \cite{Riechert_SST_2002} as well as devices emitting at 1.5 $\mu$m and beyond, with characteristics comparable to conventional InP-based devices. Overall good characteristics have been demonstrated by dilute nitride lasers, but the advantages of using dilute nitrides e.g. in VCSEL structures have not been sufficiently pronounced to displace the well-tested incumbent technology based on InP. We note that the lack of uptake of dilute nitride based lasers was at least in-part driven by economic uncertainty in the telecomms industry during the mid-2000s where some of the most successful commercial producers of devices left the sector. This led to a large loss of momentum in the development of the technology. However, dilute nitrides are making a resurgence, for example, as the 1 eV junction in multi-junction solar cells which currently hold the efficiency record of 43.5\% \cite{Optics_News,Jones_patent_2011} and for use in III-V alloys for direct integration of lasers with silicon \cite{Liebich_APL_2011}, both of which are proving to be promising applications of this interesting class of materials.

The measured threshold current in dilute nitride devices includes a large Auger contribution to the total current, similar to that found in conventional InP-based devices. Combined with the persistence of intervalence band absorption in dilute nitride lasers, this can lead to a strong temperature dependence of the threshold current and optical emission characteristics, as is also the case for InP-based devices. Considerable benefit would be obtained by eliminating these loss mechanisms from telecomm devices. We have shown that a very large spin-orbit-splitting energy, $\Delta_{\scalebox{0.7}{\rm{SO}}}$, can be achieved in GaBiAs alloys, with $\Delta_{\scalebox{0.7}{\rm{SO}}}$ exceeding the energy gap $E_{g}$ for Bi compositions above 10\% and emission wavelengths around 1.5 $\mu$m. The growth of GaBiNAs or of GaInBiAs on InP can extend this condition to longer wavelengths. We have described that achieving $\Delta_{\scalebox{0.7}{\rm{SO}}} > E_{g}$ should suppress inter-valence band absorption and the dominant Auger recombination loss mechanism both in telecomm and in mid-infrared lasers. We argue that the introduction of dilute bismide alloys should therefore finally allow a route to achieve efficient temperature-stable lasers with significantly reduced power consumption at telecomm and longer wavelengths, offering considerable benefits for a wide range of applications.  

%%%%%%%%%%%%%%%%%%%%%
%% Acknowledgments %%
%%%%%%%%%%%%%%%%%%%%%

\ack

We are grateful to many colleagues with whom we have had the pleasure to interact on dilute nitride and bismide alloys, and particularly thank Alf Adams, Henning Riechert and Stanko Tomi\'{c} for their major contributions to the work described here. This work on dilute bismide alloys is supported by the European Union Seventh Framework Programme (BIANCHO; FP7-257974), Science Foundation Ireland (10/IN.1/I299), the Irish Research Council for Science, Engineering and Technology (RS/2010/2766) and the UK Engineering and Physical Sciences Research Council (EP/H005587/1, EP/G064725/1). MU acknowledges computational resources from the National Science Foundation (NSF) funded Network for Computational Nanotechnology (NCN) through http://nanohub.org.

%%%%%%%%%%%%%%%%
%% References %%
%%%%%%%%%%%%%%%%

\section*{References}

\bibliographystyle{iopart-num}
%\bibliography{review-11.05.12}
\providecommand{\newblock}{}

\end{document}